%% file: main.tex
\documentclass[lettersize,journal]{IEEEtran}
\usepackage{amsmath,amsfonts}
\usepackage{algorithmic}
\usepackage{algorithm}
\usepackage{array}
\usepackage[caption=false,font=normalsize,labelfont=sf,textfont=sf]{subfig}
\usepackage{textcomp}
\usepackage{stfloats}
\usepackage{url}
\usepackage{verbatim}
\usepackage{graphicx}
\usepackage{cite}

\usepackage{svg}
\usepackage{svg-extract}
\graphicspath{{svg-inkscape/}}

\usepackage{booktabs}
\usepackage{tabularx}

\usepackage{bm}

\usepackage{layouts}

\usepackage{pifont}%
\newcommand{\xmark}{\ding{55}}%

\usepackage{siunitx}

\newcounter{footnotehewing}
\newcounter{footnotezeroorder}

\newcounter{footnotecrscode}
\newcounter{footnotecrsdata}
\newcounter{footnotelacados}

\usepackage{xcolor}

\usepackage{doi}
\usepackage[capitalise]{cleveref}

\usepackage{placeins}

\usepackage{makecell}

\usepackage{array}

\input{macros.tex}

\usepackage{pythonhighlight}

\begin{document}

\title{
    L4acados: 
    Learning-based models for acados, \\
    applied to 
    Gaussian process-based 
    predictive control
}

\author{%
    Amon~Lahr$^{1,\star}$, %
    Joshua~N\"af$^{1}$, %
    Kim~P.~Wabersich$^2$, %
    Jonathan~Frey$^3$,  %
    Pascal Siehl$^2$,\\
    Andrea~Carron$^1$, %
    Moritz~Diehl$^3$,
    Melanie~N.~Zeilinger$^1$%
    \thanks{$^\star$Corresponding author.}
    \thanks{$^{1}$%
    ETH Zurich, Zurich, Switzerland. E-mail correspondence to  {\tt \{amlahr,naefjo,carrona,mzeilinger\}@ethz.ch}.}
    \thanks{$^{2}$Robert Bosch GmbH, Corporate Research, Stuttgart, Germany. E-mail: {\tt \{kimpeter.wabersich,pascal.siehl\}@de.bosch.com}.
    }
    \thanks{$^{3}$Department of Microsystems Engineering (IMTEK) and Department of Mathematics, University of Freiburg, 
    Freiburg, Germany. E-mail: {\tt \{jonathan.frey,moritz.diehl\}@imtek.uni-freiburg.de}}
    \thanks{This work was supported by the European Union's Horizon 2020 research and innovation programme, Marie~Sk\l{}odowska-Curie grant agreement No. 953348,~\mbox{ELO-X}, and by DFG via 424107692 and 504452366 (SPP 2364).
    } %
    \thanks{\textcopyright~2025 IEEE. Personal use of this material is permitted.  Permission from IEEE must be obtained for all other uses, in any current or future media, including reprinting/republishing this material for advertising or promotional purposes, creating new collective works, for resale or redistribution to servers or lists, or reuse of any copyrighted component of this work in other works.}
}

\markboth{Journal of \LaTeX\ Class Files,~Vol.~14, No.~8, August~2021}%
{Lahr \MakeLowercase{\textit{et al.}}: 
L4acados: Learning-based control with acados
}

\maketitle

\begin{abstract}
Incorporating learning-based 
models, 
such as artificial neural networks or Gaussian processes, 
into model predictive control~(MPC) strategies
can significantly improve control performance and online adaptation capabilities for real-world applications. 
Still,
enabling state-of-the-art implementations of learning-based models
for MPC is complicated by the challenge of interfacing
machine learning frameworks 
with 
real-time optimal control
software.
This work aims at filling this gap by
incorporating 
external sensitivities in 
sequential quadratic programming
solvers for 
nonlinear 
optimal control.
To this end,
we provide
\texttt{L4acados}, 
a general framework for incorporating 
\texttt{Python}-based
dynamics 
models in 
the real-time optimal control software
\texttt{acados}.
By 
computing
external
sensitivities 
via 
a user-defined \texttt{Python} module,
\texttt{L4acados}
enables the implementation
of 
MPC controllers 
with learning-based residual models 
in \texttt{acados},
while 
supporting
parallelization 
of sensitivity computations
when preparing the quadratic subproblems.
We demonstrate 
significant speed-ups
and superior scaling properties
of \texttt{L4acados}
compared to 
available software using a neural-network-based control example.
Last, 
we 
provide an efficient and modular real-time implementation 
of Gaussian process-based MPC using \texttt{L4acados},
which is
applied to two hardware examples:
autonomous miniature racing,
as well as 
motion control of a full-scale autonomous vehicle for 
an 
ISO
lane change maneuver.
\begin{center}
    Code: \url{https://github.com/IntelligentControlSystems/l4acados} \\
    Video: \url{https://youtu.be/6tnhRnJSwW4}
\end{center}
\end{abstract}

\begin{IEEEkeywords}
Predictive control for nonlinear systems, control software, machine learning.
\end{IEEEkeywords}

\section{Introduction}

\IEEEPARstart{A}{ugmenting}
physics-based prediction models with learning-based 
components,
such as 
neural networks 
or Gaussian processes,
has shown to be 
a data-efficient
way 
to improve 
their
prediction accuracy
at reduced modeling effort.
Leveraging the improved prediction capabilities in model predictive control~(MPC) architectures~\cite{hewing_learning-based_2020},
the gray-box modeling approach has shown to be
highly effective
for 
challenging applications such as 
autonomous driving~\cite{kabzan_learning-based_2019,hewing_cautious_2020,spielberg_neural_2022}, 
drone control~\cite{torrente_data-driven_2021,chee_knode-mpc_2022,saviolo_physics-inspired_2022,li_nonlinear_2023,chee_flying_2024}
outdoor mobile robots~\cite{ostafew_learning-based_2014,ostafew_robust_2016} 
and 
other robotic applications~\cite{carron_data-driven_2019}.

\begin{figure}
    \includegraphics[width=\columnwidth]{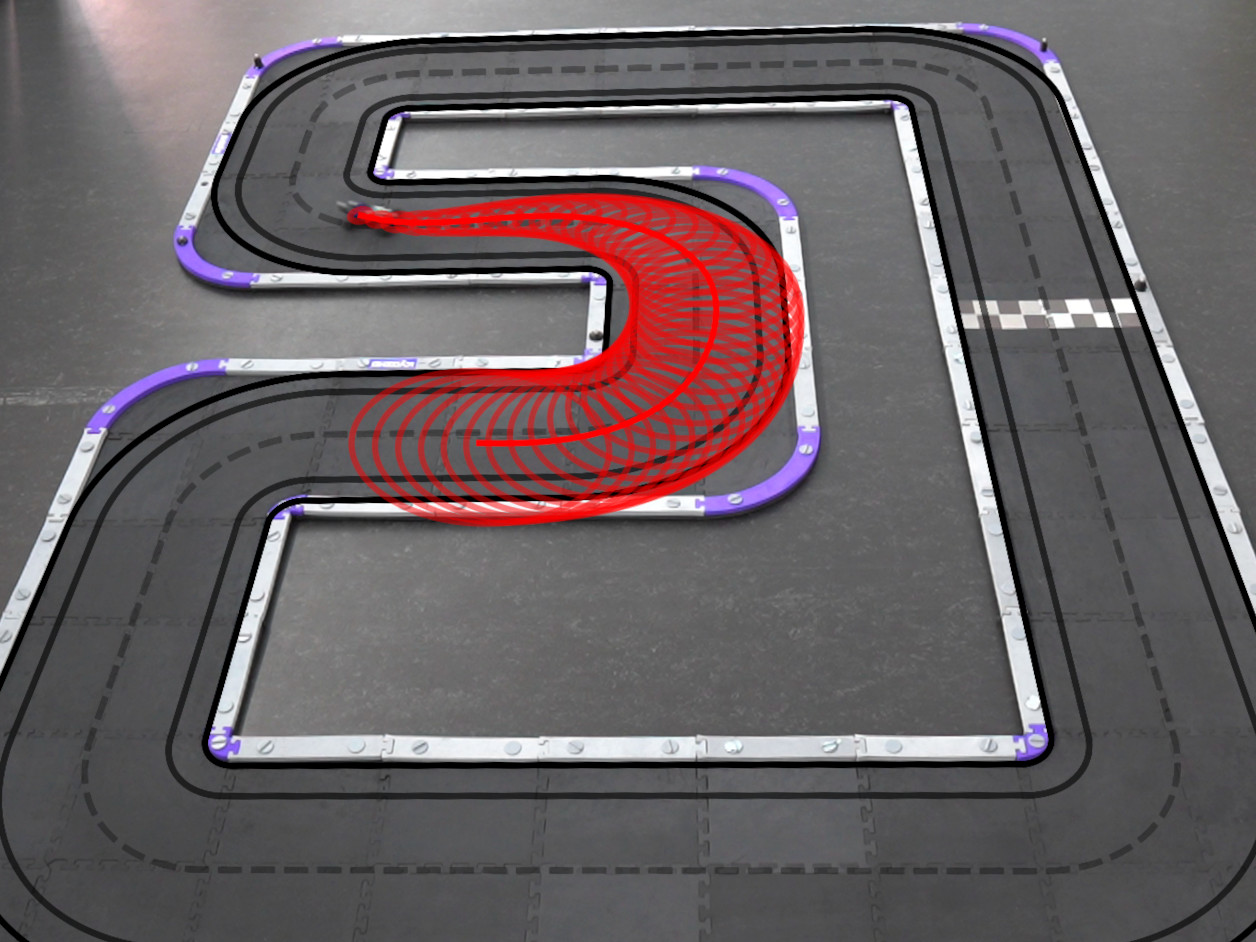}
    \caption{Real-time 
    Gaussian process-based MPC~(GP-MPC) 
    implementation with \texttt{L4acados} applied to autonomous miniature racing; uncertainty-aware predictions of the open-loop state trajectory are shown in red. 
    \texttt{L4acados} 
    enables
    efficient and parallelized external sensitivity computations of 
    general learning-based dynamics models.
    The modular and open-source GP-MPC implementation supports 
    arbitrary 
    \texttt{GPyTorch} 
    GP models,
    fast covariance propagation,
    and various data processing strategies for online learning.
    }
    \label{fig:crs_racing}
    \vspace{-1ex}
\end{figure}

MPC determines the control input in real-time by solving an 
optimization problem according to the plant dynamics model and control specifications,
requiring the 
models'
\emph{sensitivities}
i.e., 
their evaluations and higher-order 
(commonly first- and second-order) derivative information,
at each iteration step.
Integrating 
learning-based models 
into 
MPC solvers 
thus
necessitates
to 
interface 
the 
corresponding
sensitivities
with
the
embedded 
optimal control software.
While 
significant advancements have been made in 
computationally efficient
inference
and 
automatic differentiation
for learning-based models~\cite{bradbury_jax_2018,paszke_pytorch_2019,abadi_tensorflow_2015},
coupling
them with optimal control software
remains challenging.
For 
fast-sampled
MPC applications,
the combination of 
\texttt{CasADi}~\cite{andersson_casadi_2019} 
and 
\texttt{acados}~\cite{verschueren_acadosmodular_2022}
has become a popular choice,
with
\texttt{CasADi}
providing
a sparsity-exploiting automatic differentiation engine 
and 
\texttt{acados}
implementing
fast 
solvers for real-time embedded optimal control. 

In this paper, 
we 
bridge this gap
by
incorporating
external 
sensitivity 
information
through 
modification
of
model
parameters
at each 
iteration
of the 
MPC solver.
Specifically,
we provide \texttt{L4acados},
which enables 
\texttt{Python}-based
sensitivities 
for 
real-time optimal control with
\texttt{acados}.
The efficiency of the implementation 
is 
demonstrated 
for
two 
learning-based MPC implementations:
1) 
in simulation for neural network-based MPC in a benchmark against available software,
and 2) 
for Gaussian process-based MPC applied to autonomous miniature racing and 
a full-scale autonomous driving example.

\subsection*{Related work}

Existing
real-time
learning-based MPC applications
commonly resort to custom 
(re-)implementations
of the chosen 
learning-based
dynamics
model architecture
in \texttt{CasADi}~\cite{saviolo_physics-inspired_2022,yuan_safe-control-gym_2022,chee_flying_2024,gordon_introducing_2024,jiahao_online_2023,jia_tube-nmpc_2024},
lagging behind the state-of-the-art 
in terms of 
available 
methods,
computational 
efficiency,
parallelization capabilities
and 
ease of use.
An alternative approach is to 
locally approximate the learning-based model as a constant parameter~\cite{torrente_data-driven_2021,li_nonlinear_2023}
or using a second-order approximation~\cite{salzmann_real-time_2023},
reducing the number of model (Jacobian) evaluations at the expense of an approximate MPC solution.

\vfill

Available software frameworks for learning-based 
optimal control 
generally employ similar strategies
to 
incorporate 
sensitivities of learning-based 
dynamics 
models: 
For MPC with neural-network models,
\texttt{do-mpc}~\cite{fiedler_-mpc_2023} 
imports
models
adhering to the 
ONNX~\cite{the_linux_foundation_onnx_nodate} 
standard as \texttt{CasADi} models.
Similarly,
\texttt{HILO-MPC}~\cite{pohlodek_flexible_2024} supports 
artificial neural networks 
and vanilla GP regression
by
re-implementing them
within
\texttt{CasADi}.
For 
Gaussian process-based MPC~(GP-MPC),
\texttt{LbMATMPC}~\cite{picotti_lbmatmpc_2022}
re-implements
the 
GP 
posterior
mean and its Jacobian 
in \texttt{CasADi} as a vector product with precomputed weights.
In the \texttt{safe-control-gym}~\cite{yuan_safe-control-gym_2022} 
GP-MPC
implementation,
GP inference is partly manually formulated with \texttt{CasADi} symbolics and partly directly evaluated in \texttt{GPyTorch}.
A more general approach is followed by
\texttt{L4CasADi}~\cite{salzmann_learning_2024},
enabling
the integration of 
\texttt{PyTorch} models 
by compiling 
them
into the
\texttt{CasADi}
computational graph.
Yet, 
beyond the restriction to traceable \texttt{PyTorch} models,
\texttt{L4CasADi}'s batch processing capabilities are
not compatible with the 
available options for 
parallelized sensitivity computation in \texttt{acados},
complicating the efficient integration of both tools.

\vfill

In particular for Gaussian process-based MPC~\cite{hewing_cautious_2020,scampicchio_gaussian_2025},
aside from the 
coupling between efficient inference algorithms and optimal control software,
an additional difficulty is posed by the complexity of the uncertainty-aware optimal-control problem formulation.
For computational tractability,
state 
covariances 
are 
thus
largely not considered 
inside the optimization problem~\cite{ostafew_learning-based_2014,torrente_data-driven_2021,picotti_lbmatmpc_2022,panetsos_gp-based_2024},
or
heuristically fixed 
based on the state-input trajectory obtained at the
previous
sampling time of the model predictive controller~\cite{carron_data-driven_2019,kabzan_learning-based_2019,hewing_cautious_2020,zarrouki_rnmpc_2024}.
This problem has been
alleviated
by leveraging the zero-order robust optimization~(zoRO) 
algorithm,
which obtains
a suboptimal, yet feasible, 
point
of 
the 
optimal control problem~(OCP) 
at 
drastically reduced computation cost. 
The zoRO algorithm, 
initially proposed in the stochastic \cite{feng_inexact_2020} and 
later in the 
robust MPC setting \cite{zanelli_zero-order_2021},
has seen successful applications in
obstacle avoidance 
problems
using nonlinear robust MPC~\cite{gao_collision-free_2023},
as well as for friction-adaptive autonomous driving~\cite{vaskov_friction-adaptive_2022}.  
For GP-MPC,
the zoRO algorithm has been analyzed by~\cite{lahr_zero-order_2023};
similar optimization strategies have also been employed by~\cite{ostafew_robust_2016},
using a log-barrier constraint relaxation 
for control of an outdoor robot,
and by~\cite{polcz_efficient_2023},
employing a linear-parameter-varying reformulation of the GP-MPC problem.
Still, 
existing efficient and open-source implementations are tailored to their specific use-case~\cite{yuan_safe-control-gym_2022,lahr_zero-order_2023},
limiting
their 
widespread
application.

\subsection*{Contribution}

The contribution of this paper is two-fold.
First,
we 
propose a method to integrate external sensitivity information into sequential quadratic programming~(SQP) solvers for nonlinear model predictive control 
using learning-based residual 
dynamics 
models. 
To this end, we provide \texttt{L4acados} -- 
an efficient implementation thereof in the real-time optimal control software \texttt{acados}. 
The framework 
supports 
general
\texttt{Python}-callable residual 
dynamics 
models, 
Real-Time Iterations~(RTI)~\cite{diehl_real-time_2005},
as well as parallelization of the model and Jacobian evaluation during the 
RTI
preparation phase.
The computational efficiency of the proposed approach 
is demonstrated 
in a benchmark against 
available software 
for
learning-based MPC 
using a
neural-network-based 
residual
dynamics 
model.

Second,
utilizing 
\texttt{L4acados},
we provide an 
efficient, 
modular implementation
of 
Gaussian process-based MPC 
for real-time applications. 
Therefore,
we
extend
the recently developed, 
efficient implementation of the zoRO algorithm~\cite{frey_efficient_2024} 
to 
the 
zero-order GP-MPC (zoGPMPC)~\cite{lahr_zero-order_2023}
setting.
The 
fast GP-MPC 
implementation supports arbitrary 
\texttt{GPyTorch}~\cite{gardner_gpytorch_2018}
(approximate) GP models, 
as well as  
data processing strategies
for
online-learning applications,
which are demonstrated in simulation and hardware 
on the automotive miniature racing platform CRS~\cite{carron_chronos_2023}.
Last,
\texttt{L4acados} is used to implement 
a 
GP-MPC 
motion controller
for an 
ISO lane change maneuver
with a full-scale autonomous vehicle.

\section{Problem setup}

We are concerned with real-time optimal control of nonlinear
dynamical systems 
of the form
\begin{align}
    \state(t+1)  = f(\state(t),u(t)) + B_d g^{\text{tr}}(\state(t),u(t)) + w(t),
    \label{eq:true_dynamics}
\end{align}
where 
\mbox{$x(t) \in \mathbb{R}^{n_x}$} 
denotes the 
state of the system,
\mbox{$u(t) \in \mathbb{R}^{n_u}$}, 
the applied control input,
and
\mbox{$w(t) \in \mathbb{R}^{n_x}$}, 
process noise.
The ground-truth dynamics are split into
a
known, 
nominal
part of the
system dynamics,
\mbox{$f:\mathbb{R}^{n_x \times n_u} \rightarrow \mathbb{R}^{n_x}$},
and an
unknown residual,
\mbox{$g^{\text{tr}}:\mathbb{R}^{n_x \times n_u} \rightarrow \mathbb{R}^{n_g}$}.
The residual
affects
certain components of the full state 
through
the matrix 
\mbox{$B_d \in \mathbb{R}^{n_x \times n_g}$},
which is assumed to have
full column rank.
While
the
nominal model is commonly obtained by numerically integrating a continuous-time model based on first principles, 
an estimate 
\mbox{$g:\mathbb{R}^{n_x \times n_u} \rightarrow \mathbb{R}^{n_g}$}
of
the 
unknown residual
can be
obtained directly in discrete-time,
for example,
by 
training a 
learning-based 
regression model
using subsequent state measurements
of the true system,
i.e., 
\begin{align}%
    y(t) &\doteq B_d^{\dagger}(\state(t+1) - f(\state(t), u(t))) \\
    &= g^{\text{tr}}(\state(t),u(t)) + v(t). \label{eq:measmod}
\end{align}
Here, the measurements~$y(t)$
are obtained by projecting the full state onto the subspace of the residual model using the Moore-Penrose pseudo-inverse
$(\cdot)^\dagger$
of $B_d$;
the corresponding measurement noise
is given by
$v(t) = B_d^{\dagger} w(t)$.

Using the 
learning-based
model, 
a generic MPC controller 
computes the control input
for 
system~\eqref{eq:true_dynamics} 
at each sampling time as
$\inputvar(t) = \inputvar[0]^\star(x(t))$,
where $\inputvar[0]^\star(x(t))$ is the first component of the optimal input sequence
solving the following OCP:
\begin{subequations} \label{eq:OCP_nonlinear_model}
    \begin{align}
        \underset{\substack{u_0, \dots, u_{N - 1}                                        \\ \optstate[0], \dots, \optstate[N]}}{\min} \quad & \sum_{k=0}^{N-1} l(\optstate[k] , u_k) + M(\optstate[N] ) \label{eq:OCP_nonlinear_model_cost} \\ %
        \mathrm{s.t.} \quad & \optstate[0] = x(t), \label{eq:OCP_nonlinear_model_init} \\
         & \optstate[{k+1}] = f(\optstate[k], \inputvar[k]) + B_d g(\optstate[k], \inputvar[k]), \label{eq:OCP_nonlinear_model_dynamics} \\
         & 0 \geq h(\optstate[k], u_k), \label{eq:OCP_nonlinear_model_constraints} \\
         & 0 \geq h_N(\optstate[N]). \label{eq:OCP_nonlinear_model_constraints_terminal}
    \end{align}
\end{subequations}
Hereby, 
the cost~\eqref{eq:OCP_nonlinear_model_cost} to be minimized consists of 
stage cost terms
\mbox{$l:\mathbb{R}^{n_x \times n_u} \rightarrow \mathbb{R}$}
and a terminal cost 
\mbox{$M:\mathbb{R}^{n_x} \rightarrow \mathbb{R}$};
it is subject to the 
initial condition~\eqref{eq:OCP_nonlinear_model_init},
the learning-based dynamics~\eqref{eq:OCP_nonlinear_model_dynamics}
along the prediction horizon~\mbox{$k=0,\ldots,N-1$},
as well as
stage-wise
and terminal constraints~\eqref{eq:OCP_nonlinear_model_constraints} and~\eqref{eq:OCP_nonlinear_model_constraints_terminal}, respectively.

\section{L4acados: Learning-based models for acados}

Software frameworks
for
embedded 
nonlinear 
model predictive control
commonly
query
sensitivities
of the 
cost, dynamics and constraint functions in~\eqref{eq:OCP_nonlinear_model}
based on
symbolic 
expressions for the dynamics model and constraints
or a direct user interface to provide the function evaluations and sensitivities, 
see, e.g., \cite{torrisi_falcopt_2017,giftthaler_control_2018,listov_polympc_2020,chen_matmpc_2019,englert_software_2019,verschueren_acadosmodular_2022}.
In the following, 
we 
present how, 
alternatively, 
external dynamics sensitivities may be included in the SQP algorithm (\cref{sec:external_sensitivities})
by parametrizing the nonlinear program.
Afterwards,
we 
introduce
\texttt{L4acados},
an efficient implementation for external sensitivities in the real-time optimal control software \texttt{acados} (\cref{sec:l4acados_implementation}),
and 
discuss its
performance benefits
for learning-based MPC using neural networks
compared with 
alternative methods for
interfacing
external sensitivities 
(\cref{sec:experiments_l4casadi}). 

\subsection{External sensitivities for SQP solvers}
\label{sec:external_sensitivities}

Nonlinear programs of the form~\eqref{eq:OCP_nonlinear_model}
can be solved using sequential quadratic programming~\cite{boggs_sequential_1995}. 
In each SQP iteration, 
the 
quadratic subproblem
\begin{subequations} \label{eq:SQP_QP}
    \begin{align}
        \underset{%
            \substack{\Delta u_0, \dots, \Delta u_{N-1} \\ \Delta \optstate[0], \dots, \Delta \optstate[N]}
        }{\min} \quad             & \sum_{k=0}^{N-1}
        \begin{bmatrix}
            \Delta \optstate[k] \\ \Delta u_k \\ 1
        \end{bmatrix}^\top
        \begin{bmatrix}
            Q_k   & S_k & q_k \\
                  & R_k & r_k \\
            \star &     & 1
        \end{bmatrix}
        \begin{bmatrix}
            \Delta \optstate[k] \\ \Delta u_k \\ 1
        \end{bmatrix}
        \label{eq:SQP_QP_cost}
        \\
        \mathrm{s.t.} \quad \> \> & \Delta \optstate[0] = 0 \\
        & \begin{aligned}
            \Delta \optstate[{k+1}] = \> & f(\optstateHat[k],\inputvarHat[k]) 
            + B_d g(\optstateHat[k],\inputvarHat[k]) 
            - \optstateHat[{k+1}] \\&+ \hat{A}_k \Delta \optstate[k] + \hat{B}_k \Delta u_k ,
        \end{aligned}
        \label{eq:SQP_QP_dynamics} 
        \\
        & 0 \geq h(\optstateHat[k], \inputvarHat[k]) + \hat{H}^x_k \Delta \optstate[k] + \hat{H}^u_k \Delta u_k, \\
        & 0 \geq h(\optstateHat[N]) + \hat{H}^x_N \Delta \optstate[N],
    \end{align}
\end{subequations}
is solved for increments 
\mbox{$\optstateHat[k] \leftarrow \optstateHat[k] + \Delta \optstate[k]$},
\mbox{$\inputvarHat[k] \leftarrow \inputvarHat[k] + \Delta \inputvar[k]$}
to the current iterates for the states~$\optstateHat[k]$ and control inputs~$\inputvarHat[k]$.
Thereby,
$Q_k,R_k,S_k$ define the employed approximation of the Lagrangian's Hessian 
and $q_k,r_k$, the cost Jacobian at stage~$k$. 
The 
original
nonlinear equality and inequality constraints
\mbox{\eqref{eq:OCP_nonlinear_model_init}-\eqref{eq:OCP_nonlinear_model_constraints_terminal}}
are linearized, with Jacobians
\begin{align}
    &\hat{A}_k                 = \left. \frac{\partial (f + B_d g)}{\partial x} \right|_{\substack{x = \optstateHat[k] \\ u = \inputvarHat[k]}}, \quad %
    \hat{B}_k                 = \left. \frac{\partial (f + B_d g)}{\partial u} \right|_{\substack{x = \optstateHat[k] \\ u = \inputvarHat[k]}}, \label{eq:SQP_ABk} \\
    & \hat{H}^{\{x,u\}}_k = \left. \frac{\partial h}{\partial \{x,u\}} \right|_{\substack{x = \optstateHat[k]           \\ u = \inputvarHat[k]}}, \quad
    \hat{H}_N = \left. \frac{\partial h_N}{\partial x} \right|_{\substack{x = \optstateHat[k]}}. \label{eq:SQP_Hk} %
\end{align}

Utilizing the fact that
each SQP iteration does not require access to the full dynamics and constraint 
sensitivities,
but 
merely
their 
evaluations
\emph{at the current iterate},
the same SQP iterates can be obtained by
reformulating
the 
nonlinear program~\eqref{eq:OCP_nonlinear_model} 
based on the linearized dynamics~\eqref{eq:SQP_QP_dynamics}:
Instead of the original
OCP
with nonlinear dynamics,
consider the following
OCP with affine dynamics,
i.e.,
\begin{subequations} \label{eq:OCP_linear_model}
    \begin{align}
        \underset{\substack{u_0, \dots, u_{N - 1}                                        \\ \optstate[0], \dots, \optstate[N]}}{\min} \quad & \sum_{k=0}^{N-1} l(\optstate[k] , u_k) + M(\optstate[N] ) \\ %
        \mathrm{s.t.} \quad & \optstate[0] = x(t) \\
         & \optstate[{k+1}] = \hat{A}_k \optstate[k] + \hat{B}_k \inputvar[k] + \hat{c}_k, \label{eq:OCP_linear_model_dynamics} \\
         & 0 \geq h(\optstate[k], u_k),                                 \\
         & 0 \geq h_N(\optstate[N]),
    \end{align}
\end{subequations}
where $\hat{A}_k, \hat{B}_k$ and
$\hat{c}_k$
are treated as 
model parameters.
Setting
\begin{align}
    \hat{c}_k & \doteq f(\optstateHat[k],\inputvarHat[k]) + B_d g(\optstateHat[k],\inputvarHat[k]) - \hat{A}_k \optstateHat[k] - \hat{B}_k \inputvarHat[k]
\end{align}
and $\hat{A}_k, \hat{B}_k$ 
at each SQP iteration
according to~\eqref{eq:SQP_ABk}
recovers the 
original dynamics linearization~\eqref{eq:SQP_QP_dynamics} exactly.
Hence,
solving~\eqref{eq:OCP_nonlinear_model}
and~\eqref{eq:OCP_linear_model}
using
SQP 
under
the same Hessian approximation 
in~\cref{eq:SQP_QP_cost}
leads to the exact same iterates.
This is the case for 
Hessian 
approximations that do not depend on
second-order information of the dynamics, 
such as the 
generalized 
Gauss-Newton approximation.
While not implemented at the time of this writing, 
in a similar fashion, 
it is also possible to 
incorporate external sensitivity information into the cost and constraints of the 
OCP.

Generally, 
the proposed method 
offers
a simple-to-implement and effective tool to 
interface existing optimization software based on Newton-type iterations,
such as SQP and interior-point methods, cf.~\cite{nocedal_numerical_2006}, with external sensitivities.
Next, 
we provide 
an
efficient
implementation of 
external 
sensitivities
for 
\texttt{acados}~\cite{verschueren_acadosmodular_2022},
a
popular 
software package 
providing
fast OCP solvers 
and numerical integrators 
for embedded optimization. 

\subsection{\texttt{L4acados}: Learning-based models for acados}
\label{sec:l4acados_implementation}

\begin{figure*}
    \centering
    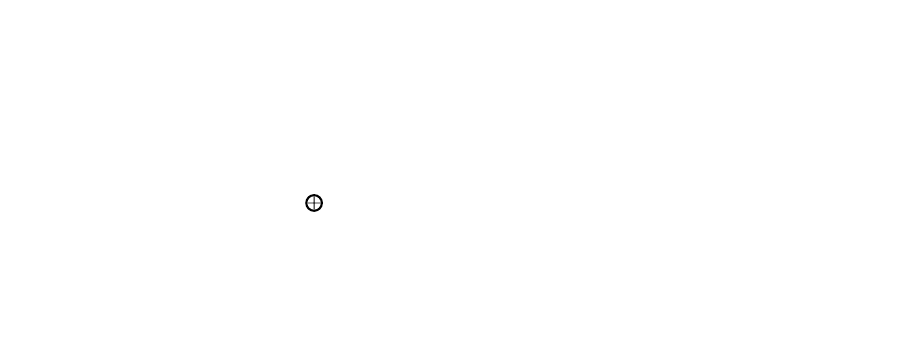
    \caption{Flow diagram of SQP iterations in \texttt{L4acados}. 
    By using
    the \texttt{acados Python} interface to \texttt{get}/\texttt{set} the model sensitivities 
    in each SQP iteration,
    \texttt{L4acados} enables learning-based models for real-time optimal control with \texttt{acados}.
    }
    \label{fig:flow_diagram_l4acados}
\end{figure*}

Within 
\texttt{acados}, 
the 
sensitivities 
in \cref{eq:SQP_ABk}
are automatically computed based on the symbolic \texttt{CasADi}~\cite{andersson_casadi_2019} expressions for the dynamics model. 
As 
\texttt{CasADi}
is in general not compatible
with 
learning-based residual models 
based on 
other automatic differentiation tools
--
with
the exception of
trace-able \texttt{PyTorch} models 
that 
can be interfaced through
\texttt{L4CasADi}~\cite{salzmann_learning_2024}
--
this complicates their use for real-time optimal control with \texttt{acados}.
To tackle this challenge, 
we present
\texttt{L4acados},
a general software framework to incorporate 
external sensitivities 
interfaced in \texttt{Python}
into 
\texttt{acados} (\cref{sec:l4acados_user_interface}).
As such, 
\texttt{L4acados} 
expands the set of 
supported \texttt{acados} models,
i.e., \texttt{CasADi} and generic \texttt{C/C++} functions, 
with \texttt{Python}-callable functions,
including
learning-based models from state-of-the-art machine learning packages and
\texttt{Python}-interfaced black-box simulators (\cref{sec:l4acados_applications}).
Computationally, 
\texttt{L4acados} 
enables
the efficient inclusion of such models by 
using \emph{batched sensitivity evaluations} (\cref{sec:l4acados_batched_sensitivities}),
which we demonstrate to be highly beneficial for
increasing horizon lengths~$N$,
larger model complexities, or
easy-to-parallelize model evaluations~(\cref{sec:experiments_l4casadi}).

\subsubsection{User interface}
\label{sec:l4acados_user_interface}

To enable 
efficient treatment of learning-based models 
in
\texttt{acados}, 
\texttt{L4acados} 
requires two user inputs (see~\cref{fig:flow_diagram_l4acados}): 
an \texttt{AcadosOcp} 
defining
the OCP formulation~\eqref{eq:OCP_nonlinear_model} for the nominal model~$f$,
and
a 
\texttt{ResidualModel}
for the 
external 
residual model~$g$.
Automatic conversion of the nonlinear OCP is performed 
by replacing the original 
model
with 
the linearized discrete-time 
dynamics~\eqref{eq:OCP_linear_model_dynamics}. 
Thereby,
the 
nominal dynamics 
are
extracted as an \texttt{AcadosSimSolver} object, 
which 
efficiently integrates the 
nominal
dynamics and computes the sensitivities~\cite{frey_fast_2023}
\begin{align}
    f(\optstateHat[k],\inputvarHat[k]), && \frac{\partial f}{\partial x}(\optstateHat[k], \inputvarHat[k]), && \frac{\partial f}{\partial u}(\optstateHat[k], \inputvarHat[k]). \label{eq:nominalmodel_gradients}
\end{align}
The
residual model's value 
and Jacobians 
\begin{align}
    g(\optstateHat[k],\inputvarHat[k]), && \frac{\partial g}{\partial x}(\optstateHat[k], \inputvarHat[k]), && \frac{\partial g}{\partial u}(\optstateHat[k], \inputvarHat[k]) \label{eq:ResidualModel_gradients}
\end{align}
at the linearization points 
are supplied by a user-defined 
residual model. 
This 
retains
flexibility, 
as any \texttt{Python}-callable programs can be used for providing the quantities in~\eqref{eq:ResidualModel_gradients},
and
avoids 
cumbersome re-implementations of existing models in \texttt{CasADi},
reducing
maintenance requirements
while
integrating into the existing open-source ecosystem.

The basic syntax of \texttt{L4acados} is as follows.
First, the user defines an instance of a \texttt{ResidualModel},
returning the corresponding sensitivities 
in~\eqref{eq:ResidualModel_gradients},
evaluated on the batch
\mbox{$\texttt{y} = ((\hat{x}_0, \hat{u}_0)^\top, \ldots, (\hat{x}_{N-1}, \hat{u}_{N-1})^\top) \in \mathbb{R}^{N \times (n_x + n_u)}$}
of
linearization points at the current SQP iteration.
For common implementations, such as \texttt{PyTorch} residual models,
at the time of this writing
\texttt{L4acados} already provides the corresponding \texttt{ResidualModel} implementation.
\begin{python}
import l4acados as l4a

class MyResidualModel(l4a.ResidualModel):
    def __init__(self, ...):
        ...

    def value_and_jacobian(self, y)
        g = ...
        dgdy = ...
        return g, dgdy
\end{python}
Second, the residual model is used to instantiate a \texttt{ResidualLearningMPC} object, 
which interfaces it with a user-defined \texttt{acados\_ocp} object defining the OCP~\eqref{eq:OCP_nonlinear_model}
for the nominal model (with nominal dynamics $x_{k+1} = f(x_k, u_k)$ 
in \cref{eq:OCP_nonlinear_model_dynamics}).
\begin{python}
from acados_template import AcadosOcp

acados_ocp = AcadosOcp()
... # define acados_ocp

residual_model = MyResidualModel(...)

l4acados_solver = l4a.ResidualLearningMPC(
    ocp=acados_ocp,
    residual_model=residual_model,
)
\end{python}
The \texttt{ResidualLearningMPC} solver is designed to be interfaced in the same way as a standard \texttt{AcadosSolver} in \texttt{acados}.
With minimal \texttt{L4acados}-specific settings,
this simplifies its usage by users familiar with \texttt{acados} 
while
reducing the requirements for its own interface documentation.

\subsubsection{Parallelized sensitivity computations}
\label{sec:l4acados_batched_sensitivities}
To enable parallelization of the 
sensitivity computation across the prediction horizon,
the \pyth{ResidualModel} is 
called \emph{once} using a batch of the linearization points
obtained from the previous SQP iteration.
In particular for computationally expensive residual models,
parallelization of the 
residual model (sensitivity) evaluation
can lead to significant speed-ups,
offsetting 
overheads incurred 
when
transferring data between different devices 
for
optimization and residual model evaluation;
this is demonstrated in
\cref{sec:experiments_l4casadi}.

\subsubsection{Real-time iteration}

A popular strategy to reduce latency of the optimization-based controller is the Real-Time Iteration~(RTI)~\cite{diehl_real-time_2005},
which 
executes a single SQP iteration per sampling time,
split into two phases:
During the
preparation phase 
the sensitivity computations are performed,
while the next initial state~$x(t)$ 
is not yet available.
After 
it
has been received,
during the feedback phase, 
the prepared QP is solved and the first input is applied to the real system. 
\texttt{L4acados} supports 
the same splitting
to perform the external sensitivity computations.
Generally,
the 
approach taken in \texttt{L4acados} can be seen as 
performing (parts of) the sensitivity computations
during the RTI preparation phase outside of \texttt{acados}.
During \texttt{acados}' internal preparation phase 
(see \cref{fig:flow_diagram_l4acados}),
the remaining computations 
for setting up the quadratic subproblem
are 
executed.

\subsubsection{Promising use-cases} 
\label{sec:l4acados_applications}

Since the numerical values of $\hat{A}_k, \hat{B}_k, \hat{c}_k$
in \cref{eq:OCP_linear_model_dynamics}
are directly passed as model parameters,
\texttt{L4acados}
enables a variety of possible use cases:
\begin{itemize}
    \item Learning-based models implemented in machine learning frameworks such as 
    \texttt{PyTorch}~\cite{paszke_pytorch_2019}, \texttt{TensorFlow}~\cite{abadi_tensorflow_2015} and \texttt{JAX}~\cite{bradbury_jax_2018};
    \item Differentiable simulators interfaced in \texttt{Python}, such as \texttt{MuJoCo}~\cite{todorov_mujoco_2012} or \texttt{PyBullet}~\cite{coumans_pybullet_2016};
    \item Black-box models 
    without 
    gradient information 
    (returning no Jacobian 
    for components of the dynamics model
    or employing
    a finite-difference approximation);
    \item Custom Jacobian approximations, 
    e.g.,
    multi-level iterations~\cite{biegler_1_2007}, feasible SQP~\cite{numerow_inherently_2024}, with learning-based models; 
    \item Online model updates in \texttt{Python}.
\end{itemize}

In the following, 
we first demonstrate the use of \texttt{L4acados} 
with
\texttt{PyTorch} 
neural network models in \cref{sec:experiments_l4casadi}.
Finally, 
we deploy \texttt{L4acados} 
for Gaussian process-based MPC
using a \texttt{GPyTorch} GP implementation in 
\cref{sec:L4acados_for_GPMPC}.

\subsection{\texttt{L4CasADi} vs. \texttt{L4acados}}
\label{sec:experiments_l4casadi}

In this section, 
\texttt{L4acados} 
and
\texttt{L4CasADi}~\cite{salzmann_learning_2024}
are compared
in terms of their
computational efficiency and scaling
in 
providing learning-based sensitivities for \texttt{acados}.
\texttt{L4CasADi} integrates learning-based \texttt{PyTorch} expressions 
in the automatic differentiation framework \texttt{CasADi}
by just-in-time~(JIT) compiling them into \texttt{TorchScript} and interfacing them via \texttt{CasADi}'s 
\texttt{C/C++} interface. 
This allows for integrating the resulting \texttt{CasADi} expressions with a variety of different solver interfaces built upon \texttt{CasADi}.
For multi-layer perceptrons, ``\texttt{Naive}'' \texttt{L4CasADi} also provides the option to generate a pure \texttt{CasADi} model, eliminating additional overheads introduced by interfacing the \texttt{TorchScript} model;
for brevity, we will simply refer to this approach as \texttt{CasADi}.
In contrast, \texttt{L4acados} integrates the sensitivities of external \texttt{Python} models into \texttt{acados} on the solver-interface level. 
While this approach limits 
its applicability to \texttt{acados} in terms of solvers,
\texttt{L4acados} can interface a larger variety of external models (see \cref{sec:l4acados_applications}).
Additionally, the specialization to \texttt{acados} allows for enhancing the solver-specific computational efficiency of sensitivity evaluations in \texttt{L4acados}, as shown in the following comparison between both approaches. 

The benchmark is performed  
for a 
double-integrator system
with 
added
neural-network~(NN) dynamics%
\stepcounter{footnote}%
\setcounter{footnotelacados}{\thefootnote}%
\footnote[\thefootnotelacados]{Code: \url{https://github.com/IntelligentControlSystems/l4acados}}, 
\begin{align}
    \label{eq:integrator_dynamics}
    \begin{bmatrix}
        x_1(k+1) \\
        x_2(k+1)
    \end{bmatrix}
    = 
    \begin{bmatrix}
        x_1(t) + \Delta T x_2(t) \\
        x_2(t) + \Delta T u(t)
    \end{bmatrix}
    + 
    \begin{bmatrix}
        1 \\
        1
    \end{bmatrix}
    g(x(t))
\end{align}
where all weights
of the NN 
have been set to zero, 
i.e., 
\mbox{$g(x(t)) = \mathrm{NN}(x(t)) \equiv 0$}. 
As 
similarly 
done in \cite{salzmann_real-time_2023},
this allows one to scale the number of hidden layers, the ``depth''~$d$, %
and number of neurons per hidden layer, the ``width''~$w$, %
of the multi-layer perceptron without affecting the MPC solution.
To cover different model architectures, a ``small'', a ``deep'', and a ``wide'' NN architecture is tested, with $(d, w) \in \{ (1, 256), (16, 256), (1, 1024) \}$ and
a total of $\{ 66817, 1053697, 1053697 \}$ internal model parameters, respectively.
To model the discrete-time dynamics~\eqref{eq:integrator_dynamics}
using \texttt{L4CasADi}, the nominal integrator dynamics is augmented with the neural network 
in continuous-time and then integrated inside \texttt{acados} using Euler
with time step \mbox{$\Delta T = 0.05$};
using \texttt{L4acados}, the continuous-time nominal dynamics is integrated using the same integrator
and then augmented using the neural network dynamics. 
This way, both approaches solve the same quadratic subproblems at each SQP iteration, 
leading to the same iterates up to numerical accuracy.

\begin{figure*}
    \centering
    \includegraphics[trim=10 0 0 0, clip, scale=1.0]{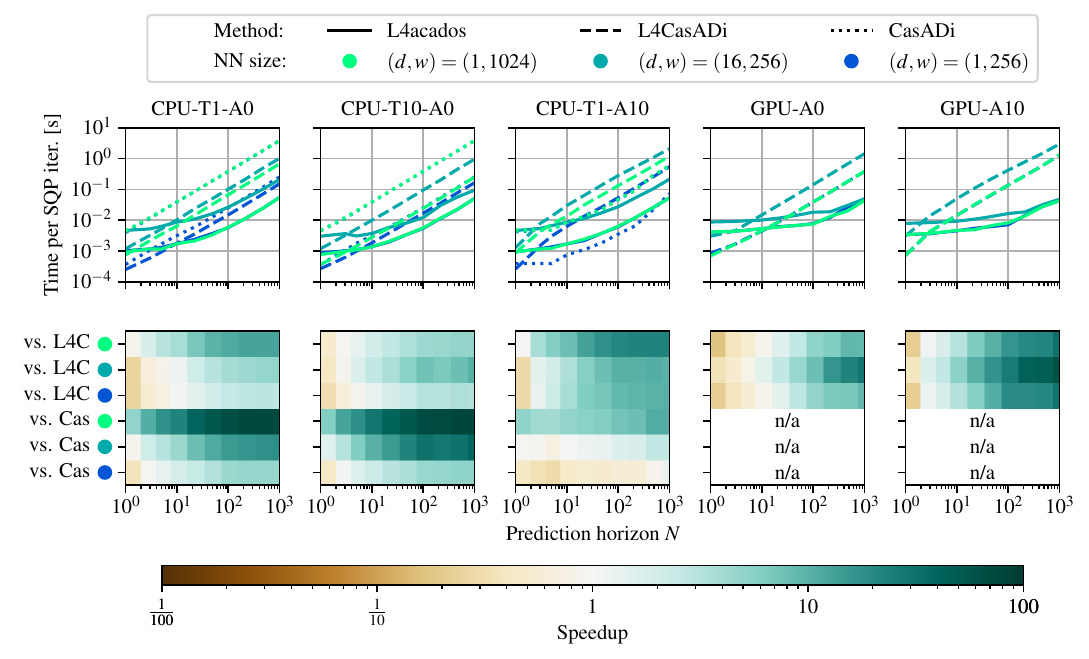}%
    \caption{Comparison of 
    computation times for sensitivity evaluation
    for neural-network residual models
    between \texttt{L4acados} and 
    \texttt{acados} with 
    \texttt{L4CasADi} 
    or \texttt{CasADi} (\texttt{Naive} \texttt{L4CasADi}).
    Already for small horizon lengths and moderate learning-based-model complexities, \texttt{L4acados} shows 
    considerable
    speedups 
    compared with
    using \texttt{acados} with \texttt{L4CasADi} or \texttt{CasADi}.
    }
    \label{fig:l4casadi_comparison}
\end{figure*}

In \cref{fig:l4casadi_comparison}, 
computation times for sensitivity evaluations in \texttt{acados},
i.e., for the QP \emph{preparation phase},
are compared between
\texttt{L4acados},
\texttt{L4CasADi}
and 
\texttt{CasADi},
where the latter is implemented via
\texttt{Naive L4CasADi}.
All three 
approaches are compared 
in terms of their 
computation time and scaling 
(top row),
as well as the achieved speedup of \texttt{L4acados} 
with respect to \texttt{L4CasADi} and \texttt{CasADi} (bottom row),
for an 
increasing 
prediction horizon~$N$,
NN 
complexity (indicated by the size%
~$(w,d)$
of the 
multi-layer perceptron),
and parallelization configurations ``$\text{DEV}$-A$Y$''. 
Here,
\mbox{$\text{DEV} \in \{ \text{CPU-}X, \text{GPU} \}$} 
indicates whether the 
NN
is evaluated on the CPU or GPU (with CUDA), respectively.
In case of CPU parallelization,
$X$
denotes
the number of respective CPU cores
enabled for \texttt{PyTorch}.
Since \texttt{acados} 
also 
supports CPU-parallelized sensitivity evaluations using OpenMP,
the 
corresponding number~$Y$ of CPU cores used is varied as well, 
where \mbox{$Y=0$} indicates that \texttt{acados} is compiled without OpenMP parallelization. 
The following experiments have been performed on an
\mbox{Intel~i9-7940X~CPU~@~3.10GHz} 
and a \mbox{NVIDIA GeForce RTX 2080 Ti GPU}.

Overall, 
it becomes evident that 
\texttt{L4acados} 
achieves noticeable speedups 
compared with \texttt{L4CasADi} and \texttt{CasADi}
across 
most
considered parallelization configurations
for typically-short prediction horizons $N \geq 10$.
In comparison with
\texttt{L4CasADi},
the speedup is most pronounced for 
the \mbox{``-A10''}~parallelization scenarios;
in comparison with 
\texttt{CasADi}, 
for
both ``large'',
in particular the ``wide'', NN configurations.
In contrast, 
performing the
sensitivity evaluations
with
\texttt{L4CasADi} 
tends to be faster
for prediction horizons~$N < 10$,
while \texttt{CasADi} is most effective for the ``small'' NN configuration
in conjunction with \texttt{acados}' OpenMP parallelization option. 
In the following, we discuss these differences in detail 
and 
attribute them to
method-specific implementation details. 

\subsubsection*{Parallelized model evaluations}

Executing all sensitivity computations 
on a single CPU core,
the ``CPU-T1-A0'' scenario highlights 
lower 
computation times for NN inference with \texttt{PyTorch},
for both \texttt{L4CasADi} and \texttt{L4acados} compared with \texttt{CasADi},
which could be attributed to efficient vectorization of layer-wise NN evaluations in \texttt{PyTorch}.
In particular, 
\texttt{CasADi}
exhibits
strikingly similar computation times 
for 
both ``large'' models,
whose matching number of parameters can be seen as a proxy for the 
number of atomic operations for automatic differentiation.
In contrast, 
the \texttt{PyTorch} implementation
of 
the ``wide'' NN architecture 
in \texttt{L4CasADi} and \texttt{L4acados}
leads to lower 
inference
times than the ``deep'' one
enforcing sequential layer-by-layer computations.

\subsubsection*{Batched sensitivity computations}
By batching the 
sensitivity evaluations in \texttt{Python},
\texttt{L4acados}
performs a single function call for the residual model's sensitivities
instead of~$N$ separate calls for each stage of the OCP.
This avoids computational overheads and,
in case of GPU parallelization, reduces the number of memory transfers between devices.
Towards the largest horizon length tested, $N=1000$,
the achieved speedup converges to a constant factor due to the limited amount of workers available for parallelization,
recovering the linear scaling observed for \texttt{L4CasADi} and \texttt{CasADi} in the \mbox{``-A0''}~parallelization scenarios.
The benefit of parallelizing sensitivity computations along the prediction horizon
is also evident for \texttt{CasADi} with \texttt{acados}' OpenMP parallelization in the ``CPU-T1-A10'' scenario, which shows a similar sublinear scaling compared to \texttt{L4acados}.
For \texttt{L4CasADi}, the \mbox{``-A10''} scenarios with \texttt{acados}' OpenMP parallelization generally resulted in larger runtimes compared to compiling \texttt{acados} without OpenMP parallelization, which could be due to interference with OpenMP parallelization in \texttt{PyTorch}%
\footnote{In particular, using both \texttt{L4CasADi} and \texttt{acados} with OpenMP parallelization, i.e., ``CPU-T10-A10'', led to considerably slower computation times as well as leaked memory at every solver call,
and is thus not shown here.}%
. 
While implementing batch-wise sensitivity evaluations in \texttt{acados} in principle could lead to similar speedups when using \texttt{L4CasADi},
this feature is difficult to realize in \texttt{acados} due to its stage-wise-modular architecture geared towards dynamic optimization problems.

\subsubsection*{Just-in-time~(JIT) compilation}

While in \texttt{L4CasADi},
the \texttt{PyTorch} model is just-in-time~(JIT) compiled to \texttt{TorchScript} at initialization, 
the current implementation of \texttt{L4acados}
does not employ JIT-compilation and uses the interpreted,
yet heavily-optimized, \texttt{PyTorch} implementation. 
In particular for small prediction horizons~$N < 10$,
this
explains in part the difference in computation times between \texttt{L4CasADi}
and \texttt{L4acados},
and highlights potential for future improvement of \texttt{L4acados}.

In summary, 
this benchmark 
demonstrates 
that
for 
increasing horizon lengths~$N$ and model complexities,
and in particular for easy-to-parallelize model evaluations,
\texttt{L4acados} 
constitutes
a competitive tool for 
incorporating learning-based models into \texttt{acados}.

\section{Gaussian process-based MPC using L4acados}
\label{sec:L4acados_for_GPMPC}

In this section, 
we show how the modular structure of \texttt{L4acados}
allows for
an efficient implementation of 
a tailored algorithm for
Gaussian process-based MPC,
by casting the corresponding 
optimal control problem as a special case of the learning-based MPC problem~\eqref{eq:OCP_nonlinear_model}.

We begin by introducing the 
GP-MPC 
optimal control problem
formulation
(\cref{sec:GPMPC})
as well as its efficient solution using a zero-order optimization algorithm
(\cref{sec:zoGPMPC}),
followed by 
the modular implementation of the latter in \texttt{L4acados}.
The control algorithm is deployed
both 
in simulation and hardware
for real-time control on two autonomous driving platforms (see \cref{fig:hardware_platforms}):
miniature racing using CRS~\cite{carron_chronos_2023}
(\cref{sec:GPMPC_CRS})
and motion control of a full-scale 
vehicle for an 
ISO
lane change maneuver 
(\cref{sec:GPMPC_Bosch}).

\subsection{Gaussian process-based MPC formulation}
\label{sec:GPMPC}

As a special case of learning-based MPC, Gaussian process-based MPC 
estimates
the residual dynamics~$g^{\text{tr}}$, as well as the associated prediction uncertainty, using a Gaussian process 
\begin{align}
    g^{\text{tr}} \sim \mathcal{GP} \left( g, \Sigma^g \right),
\end{align}
with posterior mean 
$g: \mathbb{R}^{n_x \times n_u} \rightarrow \mathbb{R}^{n_g}$ 
and posterior covariance 
\mbox{$\Sigma^g: \mathbb{R}^{n_x \times n_u} \rightarrow \mathbb{R}^{n_g \times n_g}$},
see 
Appendix~\ref{sec:GP_basics}
on their efficient implementation.
In this setting, the process noise is assumed to be Gaussian i.i.d., i.e., 
\mbox{$w(t) \sim \mathcal{N}(0, \Sigma_w)$} 
with 
positive semi-definite 
covariance 
\mbox{$\Sigma_w \in \mathbb{R}^{n_x \times n_x}$},
which leads to 
\mbox{$v(t) \sim \mathcal{N}(0, B_d^{\dagger}\Sigma_w (B_d^{\dagger})^\top)$} 
in \cref{eq:measmod}
-- a common setup for GP-MPC as done, e.g., by~\cite{hewing_cautious_2020}. 

A major challenge for GP-MPC is 
the 
propagation of state uncertainty,
induced by the 
stochastic GP
model of the unknown function $g^{\text{tr}}$.
Commonly,
a linearization-based approximation of the stochastic dynamics in terms of its mean $\mu^x_k$ and covariance $\Sigma_k$ is employed,
which leads to the following
approximate reformulation of the GP-MPC
optimization
problem~\cite{hewing_cautious_2020}:
\begin{subequations} \label{eq:OCP_GPMPC}
    \begin{align}
        \underset{\substack{u_0, \dots, u_{N - 1}                                                                            \\ \mean[x]{0} , \dots, \mean[x]{N} , \\  \covar[x]{{0}}, \dots, \covar[x]{{N}}}}{\min} \quad & \sum_{k=0}^{N-1} l(\mean[x]{k} , u_k) + M(\mean[x]{N} ) \label{eq:OCP_GPMPC_cost} \\
        \mathrm{s.t.} \quad \quad & \mean[x]{0}  = x(t), \label{eq:OCP_GPMPC_mean_0}                                    \\
                                  & \covar[x]{0} = 0, \label{eq:OCP_GPMPC_Sigma_0}                                           \\
                                  & \mean[x]{k+1} = f(\mean[x]{k} ,u_k) 
                                  + B_d g(\mean[x]{k} ,u_k), 
                                  \label{eq:OCP_GPMPC_mean} \\
                                  & \covar[x]{{k+1}} = \Psi_k(\mean[x]{k} , u_k, \covar[x]{k}), \label{eq:OCP_GPMPC_Sigma}   \\
                                  & 0 \geq h(\mean[x]{k} , u_k) + \beta(\mean[x]{k} , u_k, \covar[x]{k}), \label{eq:OCP_GPMPC_constraints}                   \\
                                  & 0 \geq h_N(\mean[x]{N} ) + \beta_N(\mean[x]{N} , \covar[x]{N}). \label{eq:OCP_GPMPC_constraints_terminal} 
    \end{align}
\end{subequations}
Thereby, the
discrete-time covariance dynamics
are
given as
\begin{align}
        \Psi_k(\mean[x]{k} , u_k, \covar[x]{k}) = \> & A_k \covar[x]{k} A_k^\top %
        + B_d \covar[g]{} (\mean[x]{k} , u_k) B_d^\top 
        + \Sigma_w,
    \label{eq:covar_prop}
\end{align}
where
$A(\mean[x]{k} , u_k) \doteq \left. \frac{\partial}{\partial x} (f(x,u_k) + B_d g(x,u_k)) \right|_{x=\mean[x]{k}}$
denotes
the Jacobian of the nominal and GP mean dynamics with respect to the (mean) state.
Component-wise
for each \mbox{$j = 1,\ldots,n_h$},
the nominal constraints $h_j$
are tightened
by
\begin{align}
    \label{eq:GPMPC_beta_tightening}
    \beta_j(\mean[x]{k} ,u_k,\Sigma_k) \doteq \gamma_j \sqrt{C_j(\mean[x]{k} ,u_k) \Sigma_k C_j^\top(\mean[x]{k} ,u_k)},
\end{align}
where 
\mbox{$C_j(\mean[x]{k} ,u_k) = \left. \frac{\partial h_j}{\partial x} (x,u_k) \right|_{x=\mean[x]{k}}$}.
Assuming a
Gaussian
state density, 
$p_j = \Phi(\gamma_j)$
thereby corresponds to the
approximate
satisfaction probability
of the 
linearized 
one-sided 
constraint,
with
$\Phi(\cdot)$ 
denoting 
the cumulative density function of a standard normal Gaussian variable.

\subsection{Zero-order algorithm}
\label{sec:zoGPMPC}

Despite the approximations, the solution of the GP-MPC OCP~\eqref{eq:OCP_GPMPC} still remains intractable for many real-time applications:
Propagation of the state covariances~\eqref{eq:OCP_GPMPC_Sigma} inside the optimal control problem~(OCP)
introduces 
a large additional number of 
optimization variables,
which
leads 
to an unfavorable scaling of the computational complexity with respect to the state dimension;
additionally,
expensive GP computations
are to be performed, 
in particular for the gradients of the posterior covariance
and for
the Hessian of the posterior mean.
To tackle these challenges, 
it has been proposed 
in \cite{lahr_zero-order_2023}
to
employ 
the
zero-order robust optimization (zoRO) 
method \cite{feng_inexact_2020,zanelli_zero-order_2021},
which
performs 
sequential quadratic programming~(SQP)
with a tailored Jacobian approximation
to obtain a suboptimal, yet feasible, point
of the
OCP~\eqref{eq:OCP_GPMPC}
at drastically reduced computational cost.
Starting with an initial guess for the input and mean variables $\inputvarHat[k]$ and $\meanHat[x]{k}$, 
respectively,
the zoRO algorithm 
propagates
the 
covariances~\eqref{eq:OCP_GPMPC_Sigma}
based on the previous SQP iterate ($\inputvarHat[k],\meanHat[x]{k}$), 
\begin{align}
    \covarHat[x]{{k+1}} = \Psi_k(\meanHat[x]{k} , \inputvarHat[k], \covarHat[x]{k}), \quad k=0,\ldots,N-1,
    \label{eq:zoro_covariance_propagation}
\end{align} 
before obtaining the next iterate
by performing
SQP
iterations 
towards a 
Karush-Kuhn-Tucker (KKT)
point of the 
reduced-size 
OCP
\begin{subequations} \label{eq:OCP_GPMPC_zoro}
    \begin{align}
        \underset{\substack{u_0, \dots, u_{N - 1}                                                                            \\ \mean[x]{0} , \dots, \mean[x]{N}}}{\min} \quad & \sum_{k=0}^{N-1} l(\mean[x]{k} , u_k) + M(\mean[x]{N} ) \label{eq:OCP_GPMPC_zoro_cost} \\
        \mathrm{s.t.} \quad \quad & \mean[x]{0}  = x(t), \label{eq:OCP_GPMPC_zoro_mean_0}                                    \\
                                  & \mean[x]{k+1} = f(\mean[x]{k} ,u_k) 
                                  + B_d g(\mean[x]{k} ,u_k), 
                                  \label{eq:OCP_GPMPC_zoro_mean} \\
                                  & 0 \geq h(\mean[x]{k} , u_k) + \hat{\beta}_k,                    \\
                                  & 0 \geq h_N(\mean[x]{N} ) + \hat{\beta}_N.
    \end{align}
\end{subequations}
At each SQP iteration of the zoRO algorithm, 
the constraint tightenings
in \cref{eq:OCP_GPMPC_constraints,eq:OCP_GPMPC_constraints_terminal},
respectively,
are thus fixed 
--
hence, 
their Jacobians neglected
--
based on the 
iterates~$(\meanHat[x]{k},\inputvarHat[k],\hat{\Sigma}_k)$
at the previous zoRO iteration, 
i.e.,
\begin{align}
    \hat{\beta}_k \doteq \beta(\meanHat[x]{k},\inputvarHat[k],\hat{\Sigma}_k).
    \label{eq:zoro_tightening}
\end{align}
Due to the neglected Jacobian,
the zoRO algorithm obtains
a suboptimal, yet feasible, 
point
of 
the OCP at convergence.
Note that the presented version of the 
zoRO 
algorithm 
employs an additional Jacobian approximation compared to the one in~\cite{zanelli_zero-order_2021,lahr_zero-order_2023} by neglecting the entire Jacobians of the 
constraint tightenings,
as it has also been done in \cite{gao_collision-free_2023,frey_efficient_2024}.

Both nonlinear programs~\eqref{eq:OCP_GPMPC} and~\eqref{eq:OCP_GPMPC_zoro} can be expressed in the general form~\eqref{eq:OCP_nonlinear_model}, 
which allows for their implementation via \texttt{L4acados}.

\subsection{Zero-order GP-MPC implementation using \texttt{L4acados}}
\label{sec:GPMPC_implementation}

Using 
\texttt{L4acados}, 
we implement 
the zero-order GP-MPC method~\cite{lahr_zero-order_2023} 
in a modular fashion
by 
splitting it into 
    1)~a \texttt{GPyTorch}~\cite{gardner_gpytorch_2018} \pyth{ResidualModel} for efficient and parallelizable GP computations and online data processing capabilities, 
    2)~\texttt{L4acados} for interfacing the learning-based sensitivities with \texttt{acados},
    3)~an extended \texttt{acados} \emph{custom update function} implementing the zoRO algorithm. 
Therefore, we make the following software contributions.

\subsubsection{\texttt{GPyTorch} \pyth{ResidualModel}}

While the main function of the \pyth{ResidualModel} is to provide evaluations and sensitivities of the GP posterior mean, 
for the zoGPMPC implementation
it has been extended 
with additional 
features.
\subsubsection*{
GP posterior covariance%
}
To solve the quadratic subproblems of~\eqref{eq:OCP_GPMPC_zoro} in the zoGPMPC algorithm,
in addition to the GP posterior mean and its Jacobian,
evaluation of the GP posterior covariance
is required 
for
the tightenings~\eqref{eq:zoro_tightening}.
The \texttt{GPyTorch} \pyth{ResidualModel} efficiently computes the GP posterior covariance alongside with the GP posterior mean, where it can be accessed for propagating the covariances~\eqref{eq:zoro_covariance_propagation}.
Depending on the employed GP-MPC formulation,
computation of the posterior covariance can be employed at each SQP iteration
or at each sampling time,
as done in the zoGPMPC algorithm or in the heuristic implementation by~\cite{hewing_cautious_2020}, respectively.

\subsubsection*{
Input feature selection%
}
To improve data-efficiency and expressivity of the learning-based model,
in practice it can be desirable to 
reduce the size of the feature space or include additional input features beyond the states and control inputs~$(x, u)$ defined in the MPC problem~\eqref{eq:OCP_GPMPC_zoro}.
This is enabled by composing the \pyth{ResidualModel} with a general \pyth{FeatureSelector} $\phi: \mathbb{R}^{n_x} \times \mathbb{R}^{n_u} \rightarrow \mathbb{R}^{n_z}$, 
mapping the state-input pair to the desired input feature, i.e.,
$z = \phi(x, u)$.

\subsubsection*{
    Data processing strategies%
}

For online data processing, 
the \texttt{GPyTorch} \pyth{ResidualModel} can be equipped with a general 
\pyth{DataProcessingStrategy}.
The current implementation 
facilitates
the recording of data points as well as online GP model updates using a subset of data;
the latter is described in detail in Appendix~\ref{sec:GP_online_updates}. 

\subsubsection{\texttt{L4acados}}

As any general \pyth{ResidualModel} implementation,
the \texttt{GPyTorch} \pyth{ResidualModel} can be straightforwardly interfaced with the nominal  \texttt{acados} problem formulation of~\eqref{eq:OCP_GPMPC_zoro}
through \texttt{L4acados},
which sets the corresponding sensitivities accordingly during optimization.

\subsubsection{\texttt{acados} custom update function}

In between SQP iterations, the zoGPMPC algorithm propagates the state covariances according to~\eqref{eq:zoro_covariance_propagation}. 
To this end, 
we have extended the efficient zoRO implementation presented in~\cite{frey_efficient_2024}, which utilizes the high-performance linear algebra package BLASFEO~\cite{frison_blasfeo_2018}, 
to support a time-varying process noise component,
which is used to 
incorporate
evaluations of the GP posterior covariance in the covariance prediction
at every SQP iteration. 
The high-performance covariance propagation transfers the computational benefits demonstrated in~\cite{frey_efficient_2024} from the zoRO to the zoGPMPC setting.

Compared with the prototypical implementation in~\cite{lahr_zero-order_2023},
the general-purpose implementation of the zoGPMPC algorithm via \texttt{L4acados} 
simplifies its application
through 
its modular software design,
offers an easily-extendable 
set 
of functionalities in terms of input feature selection and online data processing, 
and 
exhibits 
an improved 
computational performance
due to the efficient covariance propagation.

\begin{figure}
    \centering
    \subfloat[]{%
        \includegraphics[width=0.49\columnwidth]{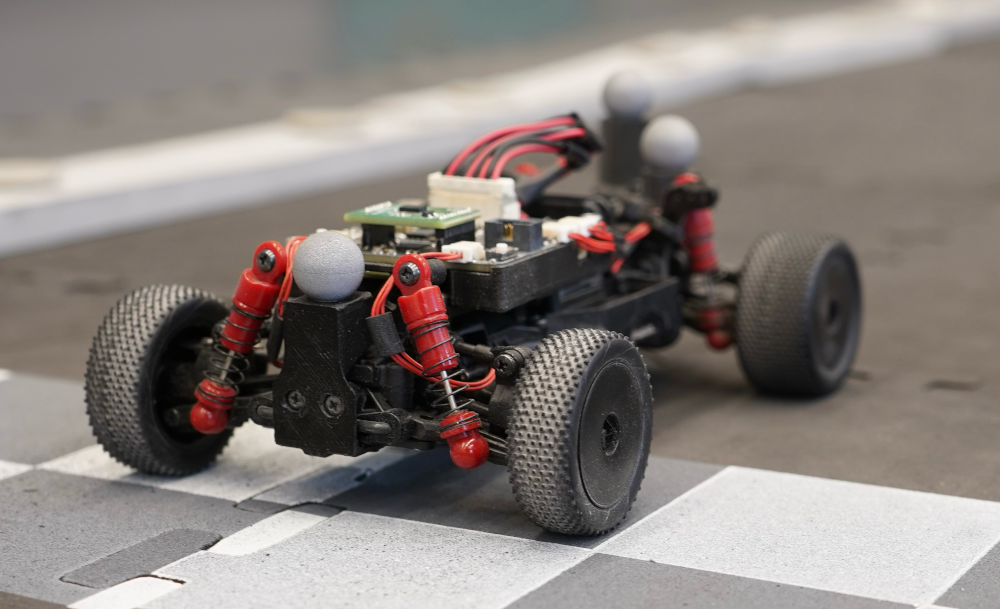}%
        \label{fig:hardware_platforms_crs}%
    }
    \hfil
    \subfloat[]{%
        \includegraphics[width=0.49\columnwidth,trim=200 100 150 215,clip]{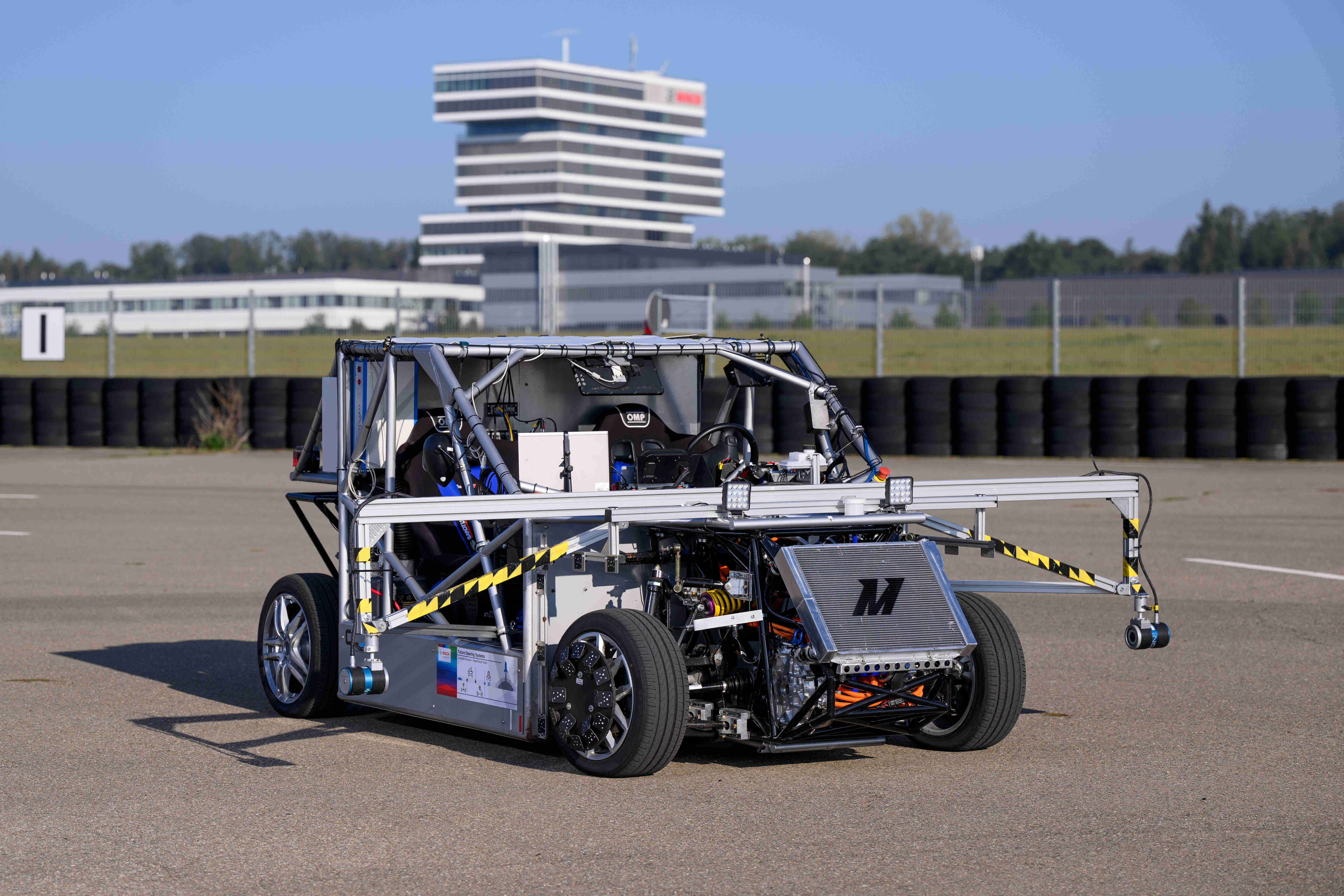}%
        \label{fig:hardware_platforms_bosch}%
    }
    \caption{Hardware platforms. Using \texttt{L4acados}, the zoGPMPC method is deployed on (a) a 1:24 miniature race car (\cref{sec:GPMPC_CRS}) and (b) a full-scale test vehicle (\cref{sec:GPMPC_Bosch}).}
    \label{fig:hardware_platforms}
\end{figure}

\subsection{Autonomous miniature racing }
\label{sec:GPMPC_CRS}

The proposed implementation is tested on the CRS robotics software platform~\cite{carron_chronos_2023}, 
which is used for autonomous racing 
in simulation%
\stepcounter{footnote}%
\setcounter{footnotecrscode}{\thefootnote}%
\footnote[\thefootnotecrscode]{Code: \url{https://gitlab.ethz.ch/ics/crs}.}
as well as on hardware%
\stepcounter{footnote}%
\setcounter{footnotecrsdata}{\thefootnote}%
\footnote[\thefootnotecrsdata]{%
Experiment data and video material: \doi{10.3929/ethz-b-000707631}. %
}%
,
using a 
custom 
1:24 miniature R/C car based on a \mbox{Mini-Z MB010} four-wheel drive chassis (see~\cref{fig:hardware_platforms_crs})
and a Qualisys motion capture system. 

\subsubsection{Car model}

The continuous-time nominal model 
of the car dynamics 
in terms of 
its mean $x$-, $y$-position and heading angle~$\psi$ in a global coordinate frame, lateral and longitudinal velocities~$v_x$ and~$v_y$, as well as yaw rate~$\omega$ in a local coordinate frame,
is 
given by the 
kinematic bicycle model 
\begin{subequations}
    \label{eq:CRS_nominal_model}
    \begin{align}
        \dot{x} &= v_x \cos(\psi) - v_y \sin(\psi) \\
        \dot{y} &= v_x \sin(\psi) + v_y \cos(\psi) \\
        \dot{\psi} &= \omega \\
        \dot{v}_x &= m^{-1} \big( F_{x,\mathrm{r}} + F_{x,\mathrm{f}} \cos(\delta) - F_{y,\mathrm{f}} \sin(\delta) \big) + v_y \omega \\
        \dot{v}_y &= m^{-1} \left( F_{y,\mathrm{r}} + F_{x,\mathrm{f}} \sin(\delta) + F_{y,\mathrm{f}} \cos(\delta) \right) - v_x \omega \\
        \dot{\omega} &= I_z^{-1} \left( - F_{y,\mathrm{r}} l_{\mathrm{r}} + F_{x,\mathrm{f}} l_{\mathrm{f}} \sin(\delta) + F_{y,\mathrm{f}} l_{\mathrm{f}} \cos(\delta)  \right)
    \end{align}
\end{subequations}
It
is 
parameterized by the mass~$m$ of the vehicle,
its inertia~$I_z$ along the $z$-axis,
and the distances of the front- and rear-wheel axle from the center of mass, 
$l_{\mathrm{f}}$ and $l_{\mathrm{r}}$, respectively. 
The
lateral tire forces at the front and rear wheel
are modeled 
using
a simplified Pacejka tire model,
\begin{align*}
    F_{y,\{\mathrm{f,r}\}} &= D_{\{\mathrm{f,r}\}} \sin \left( C_{\{\mathrm{f,r}\}} \arctan(B_{\{\mathrm{f,r}\}} \alpha_{\{\mathrm{f,r}\}} ) \right),
\end{align*}
based on the 
respective
front and rear slip angles,
\begin{align*}
    \alpha_{\{\mathrm{f,r}\}} = \arctan \left( \frac{v_y + \omega l_{\{\mathrm{f,r}\}}}{v_x} \right),
\end{align*}
cf.~\cite[Sec.~V.A.2]{carron_chronos_2023}.
The longitudinal tire forces are determined
by distributing the motor force,
\mbox{$F_{\mathrm{m}} = (C_{\mathrm{m},1} - C_{\mathrm{m},2} v_x) T$},
across the front and rear wheel according to the parameter~\mbox{$\zeta \in [0,1]$}, i.e.,
\mbox{$F_{x,\mathrm{f}} = F_{\mathrm{m}} (1 - \zeta) + F_{\mathrm{d}}$}
and
\mbox{$F_{x,\mathrm{r}} = F_{\mathrm{m}} \zeta$}.
Thereby, 
the longitudinal 
force 
includes
a polynomial friction force term
$F_\mathrm{d} = -\mathrm{sign}(v_x) \left( C_{\mathrm{d},0} + C_{\mathrm{d},1} v_x + C_{\mathrm{d},2} v_x^2 \right)$.
The state
is augmented with 
the applied torque~$T$
and steering angle~$\delta$
to model the actuator dynamics, 
and with $\theta$, the progress of the car along the track,
resulting in the state vector
\mbox{$\mean[x]{k} = (x,y,\psi,v_x,v_y,\omega,T,\delta,\theta) \in \mathbb{R}^{9}$}.
These additional states 
follow 
simple integrator
dynamics, 
whose 
corresponding velocities 
are
the control inputs~\mbox{$u_k \doteq (\dot{T}, \dot{\delta}, \dot{\theta}) \in \mathbb{R}^{3}$}.
The discrete-time nominal model~$f$ in~\cref{eq:true_dynamics} is 
defined by numerical integration using a fourth-order Runge-Kutta method 
with three stages inside the sampling interval of \mbox{$\Delta T \doteq 0.033~\mathrm{[s]}$}.
As the most severe modeling errors are considered to be in the Pacejka tire friction model and corresponding parameter estimates, 
a GP is used to learn the residual of the discrete-time velocity predictions, i.e.,
$B_d^\top \doteq \begin{bmatrix}
    0_{3 \times 3} & I_{3 \times 3} & 0_{3 \times 3}
\end{bmatrix}$.
To reduce the dimensionality of the input space, 
and the associated risk of overfitting to irrelevant features, 
the GP input features are chosen as $(v_x, v_y, \omega, T, \delta)$. 

\subsubsection{Controller setup}

The GP-MPC controller~\eqref{eq:OCP_GPMPC} is formulated 
based on
the
model predictive contouring control~(MPCC) formulation of~\cite[Eq.~(13)]{liniger_optimization-based_2015} 
with a horizon length of \mbox{$N = 40$} stages,
using a
zero terminal cost,
$M(\mean[x]{N}) = 0$,
and
a 
nonlinear-least-squares
stage cost
that 
approximates 
\begin{align*}
    l(\mean[x]{k} , u_k) \approx \int_{0}^{\Delta T} \left\| y(\mean[x]{k}, u_k) \right\|^2_{Q^y} \mathrm{d}t,
\end{align*}
through numerical integration.
Thereby,
the 
regressor
\begin{align*}
    y(\mean[x]{k}, u_k) &\doteq \left( e^c(x_k,y_k,\theta_k), e^l(x_k,y_k,\theta_k), \dot{T}_k, \dot{\delta}_k, \dot{\theta}_k, 1 \right)
\end{align*}
penalizes
the 
distance to the track's centerline 
via
approximate 
contouring and lag error terms 
\begin{align*}
    e^c(x_k, y_k, \theta_k) &\doteq \sin(\Phi(\hat{\theta}_{k})) 
    \delta x_k(\theta_k)
    - \cos(\Phi(\hat{\theta}_{k})) 
    \delta y_k(\theta_k)
    \\
    e^l(x_k, y_k, \theta_k) &\doteq -\cos(\Phi(\hat{\theta}_{k})) 
    \delta x_k(\theta_k)
    -\sin(\Phi(\hat{\theta}_{k})) 
    \delta y_k(\theta_k),
\end{align*}
as well as the control inputs $\dot{T}_k, \dot{\delta}_k$,
while rewarding progress along the track in terms of the negative penalty term on $\dot{\theta}_k$.
For computational efficiency, 
the
position~$x_{\mathrm{ref}}(\hat{\theta}_{k})$,~$y_{\mathrm{ref}}(\hat{\theta}_{k})$ and the angle~$\Phi(\hat{\theta}_k)$ of the centerline in the global coordinate system 
are evaluated at the progress variable~$\hat{\theta}_k$ of the previous MPC solution;
accordingly, the component-wise centerline
deviations
employ a linearization-based approximation
\begin{align*}
    \delta x_k (\theta_k)
    &
    \doteq 
    x_k - x_{\mathrm{ref}}(\hat{\theta}_{k}) 
    - \cos (\Phi(\hat{\theta}_k))
    \left( \theta_k - \hat{\theta}_k \right), \\
    \delta y_k (\theta_k)
    &
    \doteq 
    y_k - y_{\mathrm{ref}}(\hat{\theta}_{k})
    - \sin (\Phi(\hat{\theta}_k))
    \left( \theta_k - \hat{\theta}_k \right).
\end{align*}
Using the additional constant term in the regressor, 
the MPCC cost can be 
efficiently
expressed 
as a nonlinear-least-squares cost
with a 
positive definite
penalty matrix
\begin{align*}    
    Q^y \doteq \mathrm{diag}(Q^c,Q^l,R^T,R^\delta,S^\theta), &&
    S^\theta \doteq \begin{bmatrix}
        R^\theta & -q_\theta/2  \\ %
        -q_\theta/2 & \frac{q_\theta^2}{4 R^\theta} + \epsilon
    \end{bmatrix},
\end{align*}
given
positive 
parameters
\mbox{$Q^c, Q^l, R^T, R^\delta, q_\theta$} and any \mbox{$\epsilon > 0$}. 
Note that the value of $\epsilon$ does not influence the solution of the optimization problem, since it only affects the constant cost term.

Similar to the linearization-based approximation in the cost formulation, 
the track constraints are 
approximated by the tangents of the track boundaries 
at the centerline point based on the progress at the last MPC solution, i.e.,
via the two-sided affine constraint
\mbox{$-w_{\mathrm{track}} \leq h(\mean[x]{k}, u_k) \leq w_{\mathrm{track}}$} 
with
\begin{align}\label{eq:lateral_track_constraint}
    h(\mean[x]{k}, u_k) &= \sin(\Phi(\hat{\theta}_k)) \delta x_k (\theta_k) - \cos(\Phi(\hat{\theta}_k)) \delta y_k (\theta_k).
\end{align}
The track constraint as well as all box constraints for the state are tightened based on~\cref{eq:GPMPC_beta_tightening} with $\gamma_j = 1$. 
The detailed parameters of the box constraints for the states and control inputs can be found in the open-source implementation\footnotemark[\thefootnotecrscode].

\subsubsection{Simulation results}

To evaluate computation times in a more controlled environment for different solver configurations, 
runtime experiments are performed in 
a \texttt{ROS} simulation 
for a fixed number of $N_{\mathrm{sim}} = 3000$
simulation steps at a sampling frequency of $30\si{Hz}$, 
amounting to a 
$100 \si{s}$
time limit. 
The 
ground-truth 
car dynamics is thereby 
simulated 
using
a
set of Pacejka friction parameters obtained from system identification on the real system,
while the nominal car dynamics use values that have been slightly perturbed;
in particular, the perturbed model assumes rear-wheel drive 
($\zeta = 1$) instead of four-wheel drive ($\zeta = 0.5$) and over-estimates the 
grip, leading to unsafe driving behavior. 
\begin{figure}
    \centering
    \includegraphics[trim=0 0 0 10,clip,width=\columnwidth]{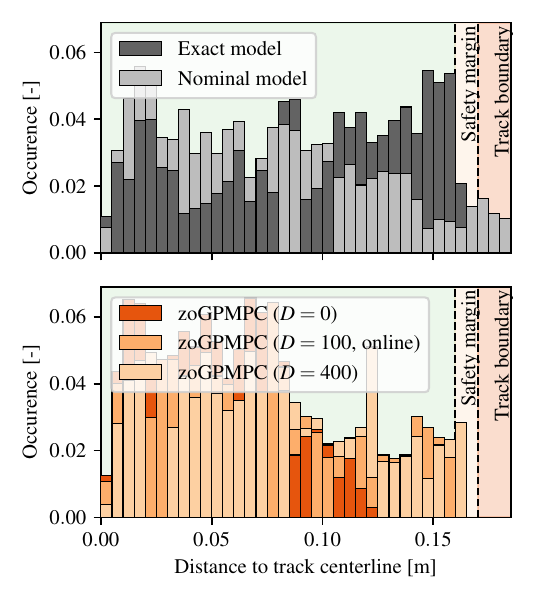}
    \caption{Distance between miniature race car and centerline in closed-loop simulations. %
    Augmented with real-world data, 
    the uncertainty-aware GP model 
    reduces
    conservatism to
    cautiously
    improve driving performance.%
    }
    \label{fig:constraint}
\end{figure}

\stepcounter{footnote}
\setcounter{footnotehewing}{\thefootnote}
\stepcounter{footnote}
\setcounter{footnotezeroorder}{\thefootnote}

\cref{fig:constraint} shows the closed-loop constraint evaluation of the track constraint for SQP-RTI controllers using different models. 
Separated by the dashed lines, the green zone denotes the (soft) constraint implemented in the controller; the yellow zone, an additional safety margin to the track boundary; the red zone, the track limits.
In the top plot, it can be seen that the controller based on the exact model operates close to the track constraint. 
While
for the exact model the car stays within the defined safety margins at all times,
for the nominal model track constraint violations (i.e. crashes) occur.
The 
small proportion 
of (soft) constraint violations using the exact model
can be attributed to approximate convergence within the real-time iterations, 
missing terminal ingredients, 
as well as 
to small control delays 
due to the real-time controller execution.
In the lower plot, the constraint evaluations for the zoGPMPC method are shown.
For $D = 0$ data points, i.e., the prior GP uncertainty, the car stays close to the centerline\footnotemark[\thefootnotezeroorder];
for $D \in \{100, 400\}$ points, much more aggressive driving behavior is observed. 
Notably, in all cases, the track constraints are respected, 
showcasing the potential for high-performance, yet uncertainty-aware,
learning-based control.

\newcolumntype{R}{>{\raggedleft\arraybackslash}p{4ex}}
\begin{table*}
    \caption{Performance comparison of different GP-MPC variants in simulation for autonomous miniature racing. 
    }
    \setlength\tabcolsep{1ex}
    \begin{center}
        \begin{tabularx}{1\textwidth}{lXXXr|RRR|RRR|RRR|c|c|c}
        \toprule
        Name & Covariance & Optimizer & GP Model & \multicolumn{1}{c|}{$D$} & \multicolumn{3}{c|}{Time, total [ms]} & \multicolumn{3}{c|}{Time, prep [ms]} & \multicolumn{3}{c|}{Time, fdbk [ms]} & Cost & Laps & Crash \\
        \midrule
        Exact model & - & SQP-RTI & - & - & 5 & 11 & 11 & 2 & 5 & 6 & 2 & 6 & 8 & 3.4 & 20.3 &  \\
        Nominal model & - & SQP-RTI & - & - & 5 & 7 & 8 & 2 & 4 & 4 & 3 & 4 & 5 & 3.4 & 20.6 & \xmark \\
        zoGPMPC & zero-order & SQP-RTI & exact & 0 & 6 & 8 & 10 & 3 & 5 & 5 & 2 & 4 & 5 & 4.5 & 14.3 &  \\
        zoGPMPC & zero-order & SQP-RTI & exact & 400 & 11 & 13 & 15 & 8 & 10 & 11 & 3 & 4 & 4 & 4.3 & 15.4 &  \\
        zoGPMPC & zero-order & SQP-RTI & exact/online & 100 & 14 & 20 & 54 & 11 & 16 & 52 & 2 & 4 & 5 & 3.9 & 17.1 &  \\
        Exact model & - & SQP & - & - & 75 & 145 & 158 & 2 & 4 & 4 & 73 & 143 & 156 & 3.3 & 21.2 &  \\
        Nominal model & - & SQP & - & - & 83 & 177 & 195 & 2 & 4 & 6 & 80 & 175 & 193 & 3.4 & 21.3 & \xmark \\
        zoGPMPC & zero-order & SQP & exact & 0 & 94 & 239 & 290 & 3 & 4 & 5 & 91 & 236 & 287 & 4.4 & 15.3 &  \\
        zoGPMPC & zero-order & SQP & exact & 400 & 171 & 357 & 374 & 7 & 11 & 11 & 164 & 349 & 366 & 4.1 & 16.4 &  \\
        zoGPMPC & zero-order & SQP & exact/online & 100 & 164 & 339 & 358 & 11 & 16 & 49 & 153 & 329 & 346 & 3.9 & 17.8 &  \\
        Cautious GPMPC\footnotemark[\thefootnotehewing] & fixed & SQP & exact & 400 & 111 & 284 & 403 & 7 & 11 & 15 & 104 & 277 & 395 & 4.2 & 16.4 &  \\
        \bottomrule
        \end{tabularx}
    \end{center}
    \label{tab:simulation_results}
\end{table*}

In \cref{tab:simulation_results}, 
a variety of solver configurations is compared:

\subsubsection*{Covariance}
\emph{zero-order} covariances denote recomputed constraint tightenings at each SQP iteration, in contrast to \emph{fixed} covariances based on the 
predicted
trajectory
at the previous sampling time\footnotemark[\thefootnotehewing];
both variants are equivalent 
for
\emph{SQP-RTI}.

\subsubsection*{Optimizer}
\emph{SQP} is used to solve the 
OCP
iteratively at each sampling time until convergence while \emph{SQP-RTI} 
performs a single iteration at each sampling time. 
For the simulations, 
an iterate 
with
all 
KKT residuals 
smaller than $\mathrm{tol} \doteq 10^{-4}$
is considered converged;
the maximum number of SQP iterations is set to $30$.
To meet sampling time requirements for all methods,
the 
simulation time is 
scaled
by 
$\alpha_{\mathrm{RTI}} \doteq 0.5$ for 
\emph{SQP-RTI}
scenarios 
and by
$\alpha_{\mathrm{SQP}} \doteq 0.015$ for 
\emph{SQP}
scenarios.

\subsubsection*{GP model}
All GP models are implemented using \texttt{GPyTorch}~\cite{gardner_gpytorch_2018}.
The \emph{exact} GP model obtains the posterior mean and covariance using an (offline) Cholesky decomposition of the kernel matrix.
Its \emph{online} implementation, 
thereby
incorporates
new data points at every sampling time, 
randomly replacing old data points once a pre-specified limit is reached, 
with recursive up- and down-dates of the kernel matrix' Cholesky factors;
see Appendix~\ref{sec:GP_online_updates}
for further details.

\subsubsection*{Time}
Total computation times (\emph{total}) are split into preparation~(\emph{prep.}) and feedback~(\emph{fdbk.}), the former including all computations that can be performed 
before sampling the next initial state, the latter, the remaining computations.
From left to right, the numbers indicate the mean, the 
99.9\% quantile, 
and the maximum computation time.
All computations for this experiment are performed on a Lenovo ThinkPad~E15~Gen2 using an Intel~i7-1165G7 processor at 2.80GHz.

\subsubsection*{Cost}
This column shows the average stage cost for the closed-loop 
trajectory
i.e., 
\mbox{$\frac{1}{N_{\mathrm{sim}}} \sum_{t=1}^{N_{\mathrm{sim}}} l(x(t), u(t))$}.

\subsubsection*{Laps}
The progress made within the 
$N_{\mathrm{sim}}$ simulation steps, in terms of the number of laps completed, is shown here.

\subsubsection*{Crash}
To compensate for delays, approximate convergence and unmodeled real-world effects, 
the (soft) track constraint 
is tightened by 
an additional safety margin (see \cref{fig:constraint}). An ``\xmark'' indicates 
violation of
the original track constraint.

Comparing the optimizer, 
it can be seen that throughout all experiments, 
SQP
performs slightly better than its 
SQP-RTI
counterpart in terms of closed-loop cost and track progress, at the expense of considerably increased computation times due to the higher number of SQP iterations. 

All 
zoGPMPC variants using SQP-RTI are  
real-time feasible;
in particular 
for the online-updating method of the exact GP model, 
computation times are 
real-time feasible
up to roughly 100 data points. 
While the maximum computation time exceeds the sampling time in exceptional cases, 
with more than 99.9\% of total computation times below 33ms, 
this does not affect the control performance in the experiments. 

Throughout all methods, 
a small closed-loop cost 
clearly correlates with a high
number of laps completed within the fixed time budget, 
highlighting the MPCC cost as an effective proxy for time-optimal racing.

\footnotetext[\thefootnotehewing]{While not exactly corresponding to the original implementation,
which employed a commercial interior-point solver,
this variant uses fixed covariances based on the last sampling time during optimization 
while solving the OCP (with correspondingly fixed tightenings) to convergence, as done in~\cite{hewing_cautious_2020}.
}

\subsubsection{Hardware experiments}

\begin{figure}
    \centering
        \includegraphics[trim=5 0 5 5, clip, width=\columnwidth]{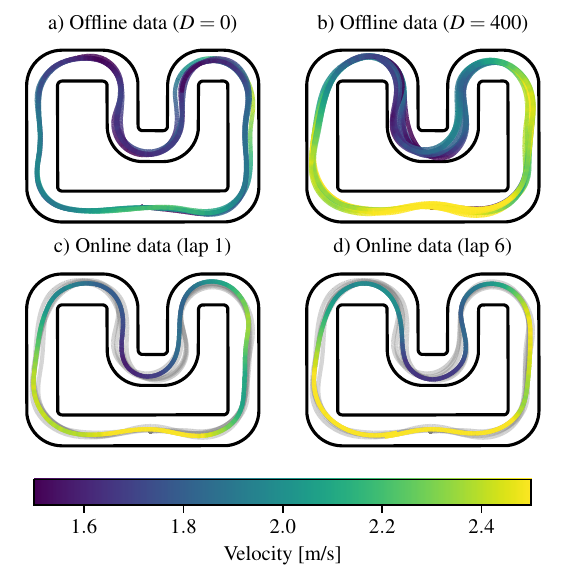}%
    \caption{Closed-loop performance comparison in miniature-racing hardware experiments (\cref{sec:GPMPC_CRS}). 
    In the offline and online learning setting, augmenting the nominal model with a GP leads to improved driving performance.
    }
    \label{fig:hw-experiments}
\end{figure}

The 
racing 
experiment has been repeated on hardware
(see \cref{fig:crs_racing}) 
using 
the
three 
\emph{SQP-RTI} 
variants of zoGPMPC in \cref{tab:simulation_results} 
with an exact GP model;
\cref{fig:hw-experiments} shows the corresponding closed-loop trajectories and car velocities%
.
As expected, the ``unconditioned'' variant with $D=0$ data points 
in Fig.~\ref{fig:hw-experiments}a) 
keeps the car closer to the centerline\footnotemark[\thefootnotezeroorder]%
\footnotetext[\thefootnotezeroorder]{%
Noteworthy, due to the 
neglected Jacobians of the covariances with respect to the states and inputs in the optimizer,
the \emph{zero-order} method, and the \emph{fixed} variant, do not take the effect of the inputs
on the size of the covariances; 
the reduction of the car velocity follows solely from the tighter track constraint, forcing the car to stay on the centerline.%
}. 
Fig.~\ref{fig:hw-experiments}b) shows the zoGPMPC method with $D = 400$ random data points recorded in the 
previous 
``unconditioned''
experiment:
The 
car drives
significantly more aggressively,
at higher speeds and taking sharper corners.
A similar driving performance is achieved using the online learning strategy with $D=100$ data points;
the transition from a more cautious to a more aggressive racing line is apparent in Fig.~\ref{fig:hw-experiments}c) and d), 
depicting the first and sixth lap of the same run, respectively.

\subsection{Motion control of full-scale vehicle}
\label{sec:GPMPC_Bosch}

To demonstrate its applicability, 
\texttt{L4acados} is used for 
safe
learning-based control of a research vehicle (\cref{fig:hardware_platforms_bosch}). 
The controller is formulated 
akin
to the autonomous-racing case in~\cref{sec:GPMPC_CRS}, 
using a 
slightly
modified system model with states
\mbox{$\mean[x]{k} = (x,y,\psi,v,\beta, \omega, \delta, \delta_{\mathrm{des}},a_x,\theta) \in \mathbb{R}^{10}$} and 
inputs \mbox{$u_k \doteq (\dot{a}_x, \dot{\delta}_{\mathrm{des}}, \dot{\theta}) \in \mathbb{R}^{3}$}. 
Here, the slip angle~$\beta$ and the effective velocity~$v$ are a different representation of $v_x$, $v_y$ in \cref{sec:GPMPC_CRS}.
The additional states $\delta_{\mathrm{des}}$ and $\delta$ represent
a first-order lag element to capture the steer-by-wire dynamics. 
The 
model equations corresponding to the modified states with respect to~\cref{eq:CRS_nominal_model}
are given by
\begin{align}
    \begin{bmatrix}
        \dot{v} \\
        \dot{\omega} \\
        \dot{\beta}\\
        \dot \delta
    \end{bmatrix}
    = 
    \begin{bmatrix}
        a_{x} \\
        \frac{1}{I_z}\left( F_{y,\mathrm{f}} \cos(\delta) l_{\mathrm{f}} - F_{y,\mathrm{r}} l_{\mathrm{r}} \right) \\
        \frac{1}{mv}(F_{y,\mathrm{f}} \cos (\beta-\delta)+F_{y,\mathrm{r}} \cos (\beta))-\omega        \\
        T_\delta^{-1}(\delta_{\mathrm{des}}-\delta)
    \end{bmatrix},
\end{align}
where 
the 
simplification \mbox{$F_{x,\{\mathrm{f,r}\}} \sin(\cdot) \approx 0$} is employed,
neglecting the longitudinal forces $F_{x,\mathrm{f}}$, $F_{x,\mathrm{r}}$ acting on the front and rear axles.
The vehicle acceleration $a_x$ results from integrating the desired acceleration rate input and 
is send to the interface provided by the vehicle platform, neglecting the underlying dynamics.
\color{black}
The model is discretized 
for $N=20$ steps
using 4th-order 
Runge-Kutta integrator 
with 
step size
$0.1~\mathrm{[s]}$. 
The motion control task is to comfortably track a given path with a reference velocity $v_{\mathrm{ref}}$, 
yielding
the modified 
cost regressor
$y(\mean[x]{k}, u_k) = ( e^c(x_k,y_k,\theta_k), e^l(x_k,y_k,\theta_k), v_k, \omega_k, \dot{a}_k, \dot{\delta}_{\mathrm{des},k}, \dot{\theta}_k, v_\mathrm{ref} )$ with diagonal 
penalty matrix~$Q_y$,
subject to
box constraints
$\beta\in[-0.17,0.17]~\mathrm{[rad]}$, $v\in[0.5,40]~\mathrm{[m/s]}$, $\delta,\delta_{\mathrm{des}}\in[0.61,0.61]~\mathrm{[rad]}$, $\dot \delta_{\mathrm{des}}\in[-0.35,0.35]~\mathrm{[rad/s]}$, $a_x\in[-5,5]~\mathrm{[m/s^2]}$, $\dot a_x\in[-4,4]~\mathrm{[m/s^3]}$.
The safety-critical constraint of staying within the track boundaries, i.e., $|e_c|\leq 1~\mathrm{[m]}$, is
formulated as in~\eqref{eq:lateral_track_constraint} with half-track width $w_{\mathrm{track}}=1$.
The GP is used to correct velocity predictions and the steering dynamics,
i.e., 
\mbox{$B_d \doteq \begin{bmatrix}
    0_{3 \times 4} & I_{4 \times 4} & 0_{3 \times 4}
\end{bmatrix}$},
with GP input features $(v, \beta, \omega, \delta, \delta_\mathrm{des}, a_x)$. 

\par
\begin{figure}
    \centering
    \includegraphics[width=\linewidth]{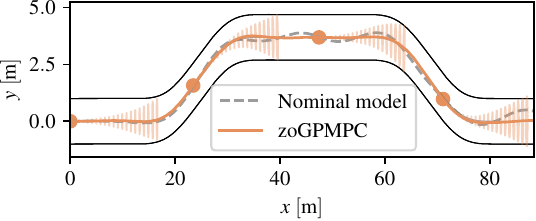} \\
    \vspace{0.1cm}

    \includegraphics[width=0.45\linewidth]{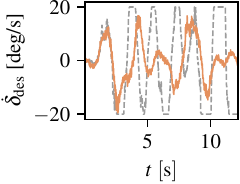}
    \hfill
    \includegraphics[width=0.47\linewidth]{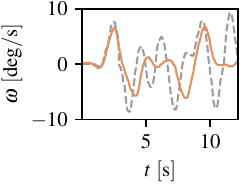}
    \caption{Double lane change maneuver of the full-scale vehicle (Figure~\ref{fig:hardware_platforms_bosch}) at $30~\si{km/h}$. The nominal MPC controller yields unsatisfactory performance; the zoGPMPC method 
    provides significant improvements. The bars indicate 95\% confidence intervals 
    that are 
    used to tighten the
    track boundary constraints.}\label{fig:experimental_results_bosch}
\end{figure}

We show the effect of including the GP 
model
via \texttt{L4acados} 
using a double lane change maneuver based on the ISO 3888-1 standard 
(without an increasing corridor)
at $30~\mathrm{km/h}$, see \cref{fig:experimental_results_bosch}.
First, a basic nominal model parameterization 
is applied, 
with \emph{SQP-RTI}. 
The control performance is unsatisfactory in terms of actuator load, comfort, and stable tracking of the desired path.
Equally spaced data points 
along the driven path are used to fit an exact GP model as described in \cref{sec:GPMPC_CRS} using basic data preprocessing. %
The experiment was repeated with the same nominal model extended by the GP correction term, 
with a 
desired probability level 
of 
$p_j \doteq 95\%$
for satisfying the safety-critical path constraint.
The result 
shows significant improvements in all performance aspects: Reduced actuation requests lead to a more comfortable driving experience, and accurate tracking of the desired path provides stable behavior. Furthermore, while the predicted confidence estimate is mostly contained within the path constraints, larger uncertainties at the end of the planning horizon act 
similar to a small terminal set constraint, further stabilizing the performance.
The computations 
performed on an HP Zbook 16 Fury G10 notebook
resulted
in total computation times 
(\mbox{mean $\pm$ standard deviation})
of 
\mbox{3.1 $\pm$ 1.4~\si{ms}} 
and 
\mbox{8.2 $\pm$ 1.4~\si{ms}} 
for the nominal MPC and zoGPMPC 
method, 
respectively.
\cref{tab:computation_time_bosch} 
displays the distribution of computation times between preparation and feedback phase\footnote{As the share of preparation and feedback time has not been measured during the full-scale experiment, the values in \cref{tab:computation_time_bosch} have been obtained by re-solving the MPC problems along the measured closed-loop trajectory. Note that this leads to a slight discrepancy between the mean total computation times observed in the actual experiment and the ones obtained by re-solving.}.
All in all, 
the computation times are
consistently
well-within the 
control cycle time
of $20\si{ms}$,
demonstrating the feasibility of real-time learning-based MPC applications using \texttt{L4acados} for a realistic autonomous driving scenario.

\begin{table*}
    \caption{Computation times for full-scale autonomous driving experiment.
    }
    \setlength\tabcolsep{1ex}
    \begin{center}
        \begin{tabularx}{1\textwidth}{XXXXR|RRR|RRR|RRR}
        \toprule
        Name & Covariance & Optimizer & GP Model & \multicolumn{1}{c|}{$D$} & \multicolumn{3}{c|}{Time, total [ms]} & \multicolumn{3}{c|}{Time, prep [ms]} & \multicolumn{3}{c}{Time, fdbk [ms]} \\
        \midrule
        Nominal model & - & SQP-RTI & - & - & 3 & 5 & 6 & 2 & 5 & 5 & 1 & 3 & 3 \\
        zoGPMPC & zero-order & SQP-RTI & exact & 136 & 6 & 11 & 11 & 5 & 10 & 10 & 1 & 1 & 1 \\
        \bottomrule
        \end{tabularx}
    \end{center}
    \label{tab:computation_time_bosch}
\end{table*}

\section{Conclusions and Outlook}

In this paper, we presented a method for incorporating external 
(learning-based) sensitivity information into optimal control solvers.
For the common use case of 
learning-based \texttt{Python} models 
in the optimal control software \texttt{acados},
we provide 
\texttt{L4acados};
the 
efficiency and applicability of the software
is demonstrated
i) in a benchmark 
against available software 
for a neural network-based MPC example
and 
ii) by
simulation and hardware experiments 
using Gaussian process-based MPC
for autonomous miniature racing and 
a full-scale autonomous vehicle prototype.
For future work, 
extensions 
to learning-based constraint and cost modules
can further increase its range of potential applications.

\section*{Acknowledgment}

The authors would like to thank
Tim Salzmann for a clarifying discussion regarding the benefits and limitations of \texttt{L4acados} compared to \texttt{L4CasADi},
Sabrina Bodmer for the CRS software support,
as well as
Lars Bartels and Alexander Hansson for their contributions to \texttt{L4acados}.

\bibliographystyle{IEEEtran}
\bibliography{IEEEabrv,bibliography_amon_do_not_edit}

\appendices

\section{Gaussian process-based MPC implementation}

In this section, we detail the implementation of the employed GP models and associated online model updates. 
For notational simplicity, in the following we assume that
the unknown ground-truth function~$g^\mathrm{tr}$ is scalar;
in the vector-valued case, the equations are then applied
component-wise.

\subsection{Gaussian process inference}
\label{sec:GP_basics}

Denote by 
$\mathcal{D} = (Z, Y)$
a data set 
of~$D$ noisy observations
$Y = \{ y_i \}_{i=1}^{D}$
at input locations
$Z = \{ z_i \}_{i=1}^{D}$,
according to
\begin{align}
    y_i = g^{\mathrm{tr}}(z_i) + w_i,
\end{align}
where the noise is i.i.d. Gaussian
with covariance~$\sigma^2$, i.e.,
\mbox{$w_i \sim \mathcal{N}(0, \sigma^2)$},
cf.~\eqref{eq:measmod} subject to the problem setup in \cref{sec:GPMPC}.
The input features $z_i = \phi(x_i, u_i)$
are determined by a user-defined feature selector~$\phi$,
see \cref{sec:GPMPC_implementation}. 
Using a positive-(semi)definite kernel function $k: \mathbb{R}^{n_z \times n_z} \rightarrow \mathbb{R}^{}$,
Gaussian process regression~\cite{rasmussen_gaussian_2006} 
determines an 
estimate
for
the unknown function~$g^{\mathrm{tr}}$
at a 
test input location~$z^\star$
and its 
associated
estimation uncertainty
as the posterior mean~$g(z^\star)$ and posterior covariance~$\Sigma^g(z^\star)$, respectively, 
with
\begin{align}
    g(z^\star) &= 
    K_{z^\star, Z} 
    \left(K_{Z,Z} + \sigma^2 I \right)^{-1}Y
    ,
    \label{eq:gp_inference_formulas_mean}
    \\
    \Sigma^g(z^\star) &= K_{z^\star,z^\star} - K_{z^\star,Z}
    \left(K_{Z,Z} + \sigma^2 I \right)^{-1}
    K_{Z,z^\star},
    \label{eq:gp_inference_formulas_covar}
\end{align}
where 
we have assumed a zero-mean prior
without loss of generality.
The kernel matrices $[K_{Z,Z}]_{ij} = k(z_i, z_j)$
as well as 
$[K_{z^\star,Z}]_{1i} = [K_{Z,z^\star}]_{i1} = k(z^\star, z_i)$
are 
defined component-wise
by evaluations of the kernel function at pairs of input locations.
The expression 
\mbox{$\left(K_{Z,Z} + \sigma^2 I \right)^{-1}Y$}
and the (incomplete) Cholesky factor 
\mbox{$LL^\top = K_{Z,Z} + \sigma^2 I$}
are cached
upon first computation in \texttt{GPyTorch}~\cite{gardner_gpytorch_2018,pleiss_constant-time_2018},
reducing the computational cost of inference 
for the same data set and hyper-parameters.
In particular, 
the latter
allows for efficient forward- and backward-substitution,
reducing 
the computational complexity for 
the linear-system solve in \cref{eq:gp_inference_formulas_covar}
from $\mathcal{O}(D^3)$ to $\mathcal{O}(D^2)$.

\subsection{Online updates in exact GP regression} 
\label{sec:GP_online_updates}

When adding a new datum with input location~$\tilde{z}$ to the data set,
the current Cholesky factor~$L^k$ of the Gram matrix can be efficiently updated in $O(D^2)$ time 
as
\begin{equation}\label{eq:cholesky_update}
    L^{k+1} =
    \begin{bmatrix}
        L^k & 0 \\
        l_2^\top & \mathrm{chol}(K_{\tilde{z},\tilde{z}} - l_2l_2^\top)
    \end{bmatrix},
\end{equation}
where $l_2$ is the solution of $L^k l_2^\top =K_{Z,\tilde{z}}$, cf.~\cite[App,~B]{osborne_bayesian_2010},~\cite[App.~A]{maiworm_online_2021}.
This is equivalent to performing an additional iteration of the row-wise Cholesky decomposition
of the augmented data set~$\mathcal{D}^{k+1}$, ordered such that the new input location is processed last,
see, e.g.,~\cite{dongarra_implementing_1984,george_parallel_1986}. 
Note that, in the considered streaming setting, 
where only one new data point is obtained at each iteration, 
the lower right block of~$L^{k+1}$ is scalar, i.e., $\mathrm{chol}(K_{\tilde{z},\tilde{z}} - l_2l_2^\top) = \sqrt{K_{\tilde{z},\tilde{z}} - l_2l_2^\top}$.

While this online update 
reduces the computation time 
of online GP inference significantly compared to a na\"ive approach, it still scales $O(D^2)$, 
exceeding the available computational budget for 
increasing data set sizes. 
To retain computational feasibility, 
upon reaching the maximum number of data points, 
we thus 
update the Cholesky factor by replacing a specific data point by the newly obtained one. 
Removing row/column $i$ from the data set can be achieved as follows
\begin{equation}
    L^{k+1} = 
    \begin{bmatrix}
        L^k_{:i, :i} & 0 \\
        L^k_{i+1:, :i} & L_+
    \end{bmatrix},
\end{equation}
where 
\begin{align*}
    L_+ \doteq \mathrm{chol}\left(L^k_{i+1:, i} (L^k_{i+1:, i})^\top + L^k_{i+1:, i+1:} (L^k_{i+1:, i+1:})^\top \right)
\end{align*}
where 
constitutes a 
rank-one update to the previous lower-right block factor. 
While significant structure could be exploited in this update, 
e.g. by applying Givens rotations~\cite[Sec.~5.1]{golub_matrix_2013},
in our preliminary experiments 
and 
for the
(fairly small) data-set sizes considered,
these approaches
fared generally worse than recomputing the factor from scratch 
due to a lacking high performance implementation. 
Our \texttt{GPyTorch} implementation of the online-updating strategy is available online%
\footnote{
\url{https://github.com/naefjo/gpytorch/blob/feature/exo-gp/gpytorch/models/exact_gp.py\#L140}.
}%
.

\section{Biography Section}

\begin{IEEEbiography}[{\includegraphics[width=1in,height=1.25in,clip,keepaspectratio]{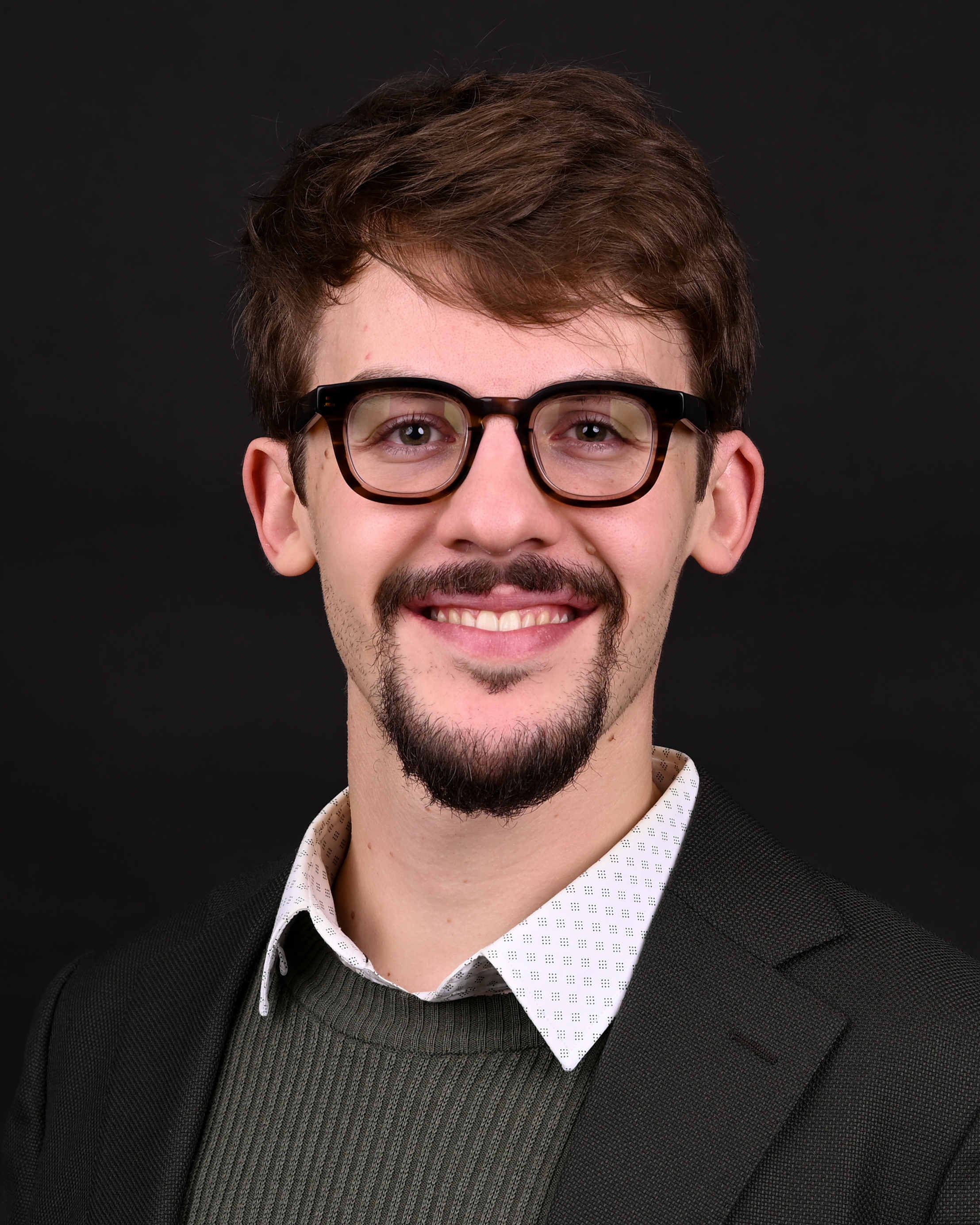}}]{Amon Lahr}
    received a bachelor's degree in engineering sciences and a master's degree in scientific computing from the Technical University of Berlin. He is currently a Ph.D. candidate 
    at ETH Zurich, Switzerland, under the supervision of Prof. Dr. Melanie Zeilinger, and a Marie
Skłodowska-Curie fellow of the European Union's innovative training network ``ELO-X''. His research interests include uncertainty-aware and learning-based control, numerical optimal control and probabilistic numerical methods.
\end{IEEEbiography}

\begin{IEEEbiography}[{\includegraphics[width=1in,height=1.25in,clip,keepaspectratio]{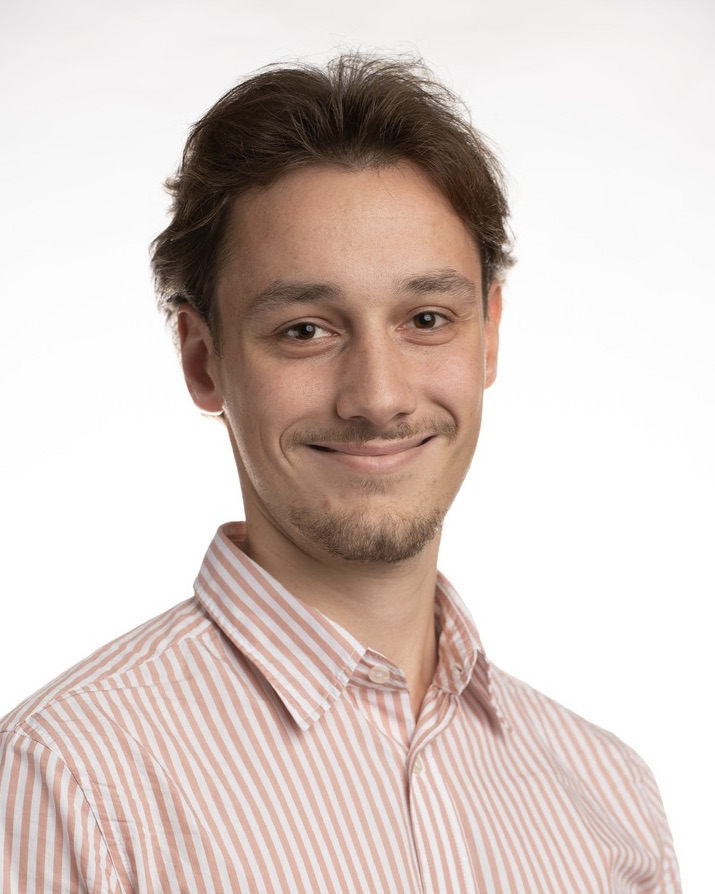}}]{Joshua N\"af}
is a Ph.D. student in the Mobile Robotics Lab at ETH Zürich under the supervision of Prof. Dr. Stefan Leutenegger. His research focus is on data-efficient policies for generalisable robotic control, particularly applied to humanoid robots.
He completed his master's degree in robotics, systems and control at ETH Zürich with distinction, focusing on the intersection of model learning techniques with predictive control methods. His master thesis was honoured with the ETH Medal.
He received his bachelor's degree in mechanical engineering from ETH Zürich, where he was involved in focus project ``Griffin", which was awarded with the best paper award by IEEE ICUAS in 2023.
\end{IEEEbiography}

\begin{IEEEbiography}[{
    \includegraphics[width=1in,height=1.25in,clip,keepaspectratio]{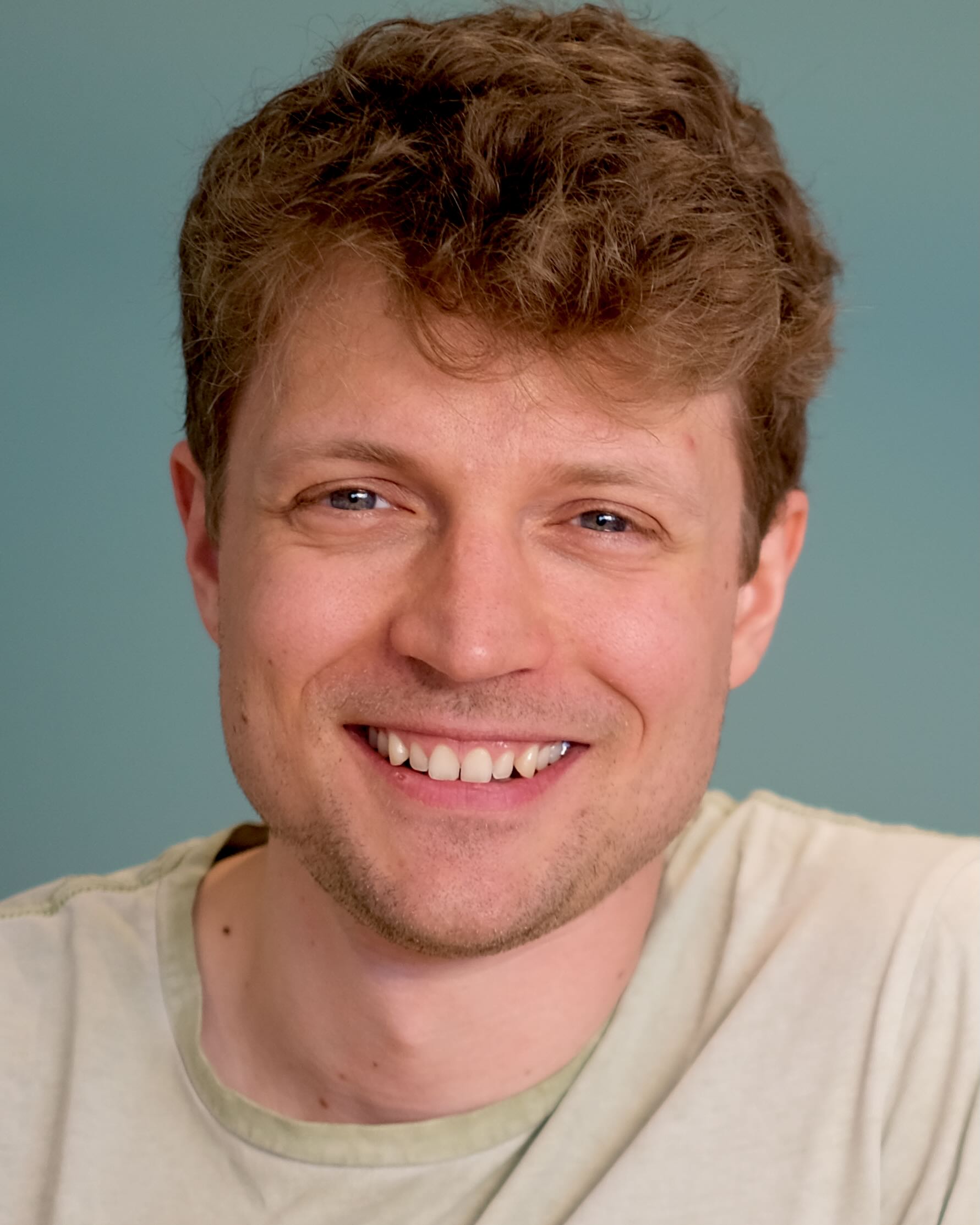}
}]{Kim P. Wabersich}
received the bachelor's and master's degrees in engineering cybernetics from the University of Stuttgart in Germany in 2015 and 2017, respectively. He received the Ph.D. degree in predictive safety mechanisms at the Institute for Dynamic Systems and Control, ETH Zurich in 2021 and continued his work as a Postdoctoral Researcher until 2022. He currently works at Bosch Research, 71272 Renningen, Germany, focusing on safety-critical systems with applications in motion control and autonomous driving. His research interests include control methods and their intersections with safe reinforcement learning.
\end{IEEEbiography}

\begin{IEEEbiography}[{
    \includegraphics[width=1in,height=1.25in,clip,keepaspectratio]{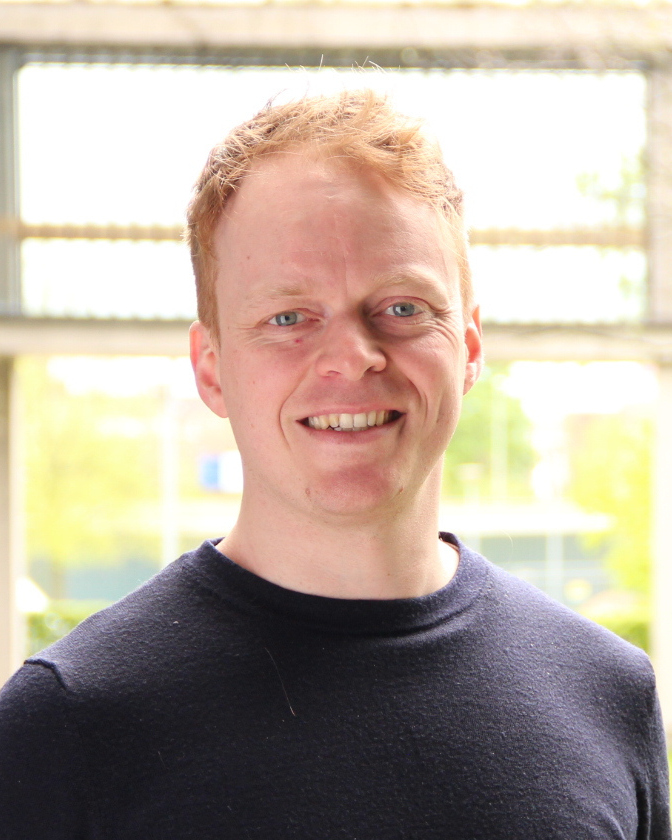}
}]{Jonathan Frey}
studied mathematics at the Technical University Ilmenau and at the University of Freiburg, where he obtained his master degree in 2019. He is pursuing a Ph.D. at the University of Freiburg on efficient algorithms, suitable formulations and differentiable solutions for optimal control under the supervision of Prof. Dr. Moritz Diehl. He is currently the maintainer of the open-source software framework “acados” which implements fast and embedded solvers for nonlinear optimal control.
\end{IEEEbiography}

\begin{IEEEbiography}[{
    \includegraphics[width=1in,height=1.25in,clip,keepaspectratio]{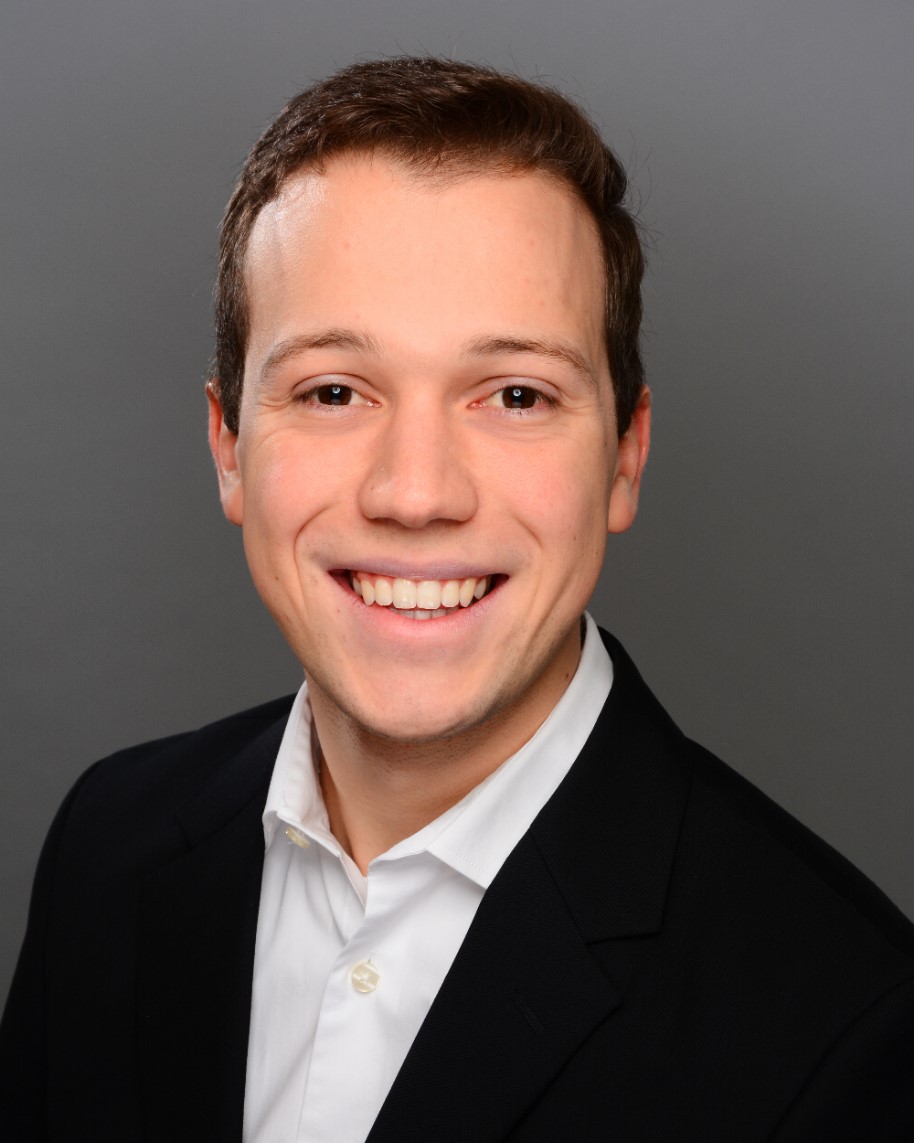}
}]{Pascal Siehl}
studied mechanical engineering at the Karlsruhe Institute of Technology (KIT) in Germany, with focus on vehicle dynamics and mechatronics and received his master's degree with distinction in 2023. He currently works as a Research Scientist at the Corporate Research department of Bosch, focusing on next-generation steering system architectures and cross-domain vehicle control strategies.
\end{IEEEbiography}

\begin{IEEEbiography}[{
    \includegraphics[width=1in,height=1.25in,clip,keepaspectratio]{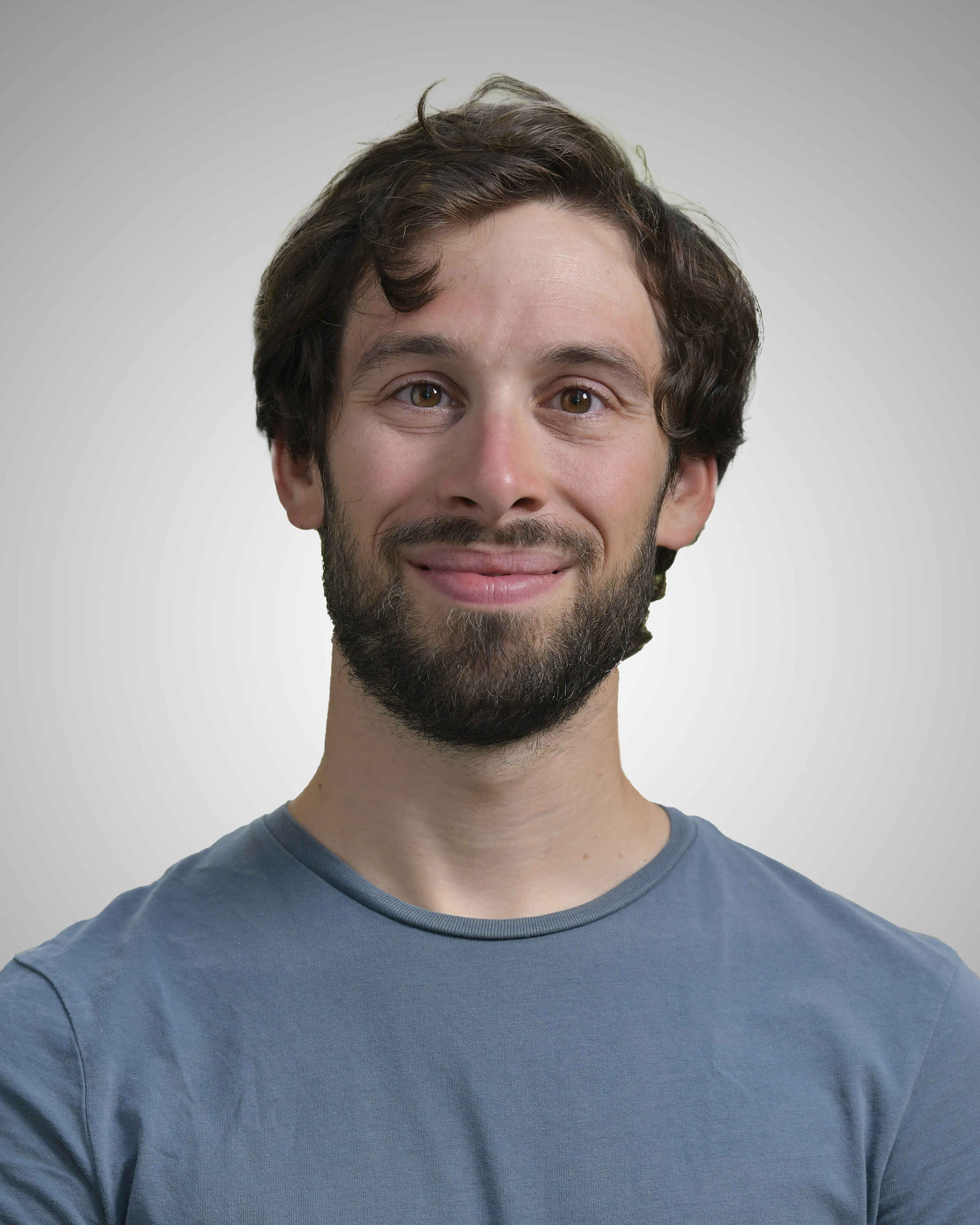}
}]{Andrea Carron}
received the bachelor's, master's, and Ph.D. degrees in control engineering from the University of Padova, Italy, in 2010, 2012, and, 2016, respectively. He is currently a Senior Lecturer with ETH Zürich. He was a Visiting Researcher with the University of California at Riverside, with Max Planck Institute in Tubingen and with the University of California at Santa Barbara, respectively. From 2016 to 2019, he was a Postdoctoral Fellow with Intelligent Control Systems Group at ETH Zürich. His research interests include safe-learning, learning-based control, multiagent systems, and robotics.
\end{IEEEbiography}

\begin{IEEEbiography}[{
    \includegraphics[width=1in,height=1.25in,clip,keepaspectratio]{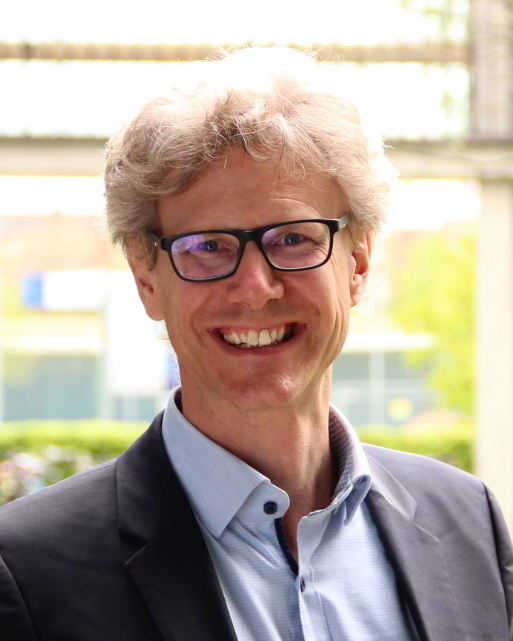}
}]{Moritz Diehl}
was born in Hamburg, Germany, in 1971. He studied physics and mathematics at Heidelberg and Cambridge University from 1993- 1999, and received his Ph.D. degree from Heidelberg University in 2001, at the Interdisciplinary Center for Scientific Computing. From 2006 to 2013, he was a professor with the Department of Electrical Engineering, KU Leuven University Belgium, and served as the Principal Investigator of KU Leuven’s Optimization in Engineering Center OPTEC. In 2013 he moved to the University of Freiburg, Germany, where he heads the Systems Control and Optimization Laboratory, in the Department of Microsystems Engineering (IMTEK), and is also affiliated to the Department of Mathematics. His research interests are in optimization and control, spanning from numerical method development to applications in different branches of engineering, with a focus on embedded and on renewable energy systems.
\end{IEEEbiography}

\begin{IEEEbiography}[{
    \includegraphics[width=1in,height=1.25in,clip,keepaspectratio]{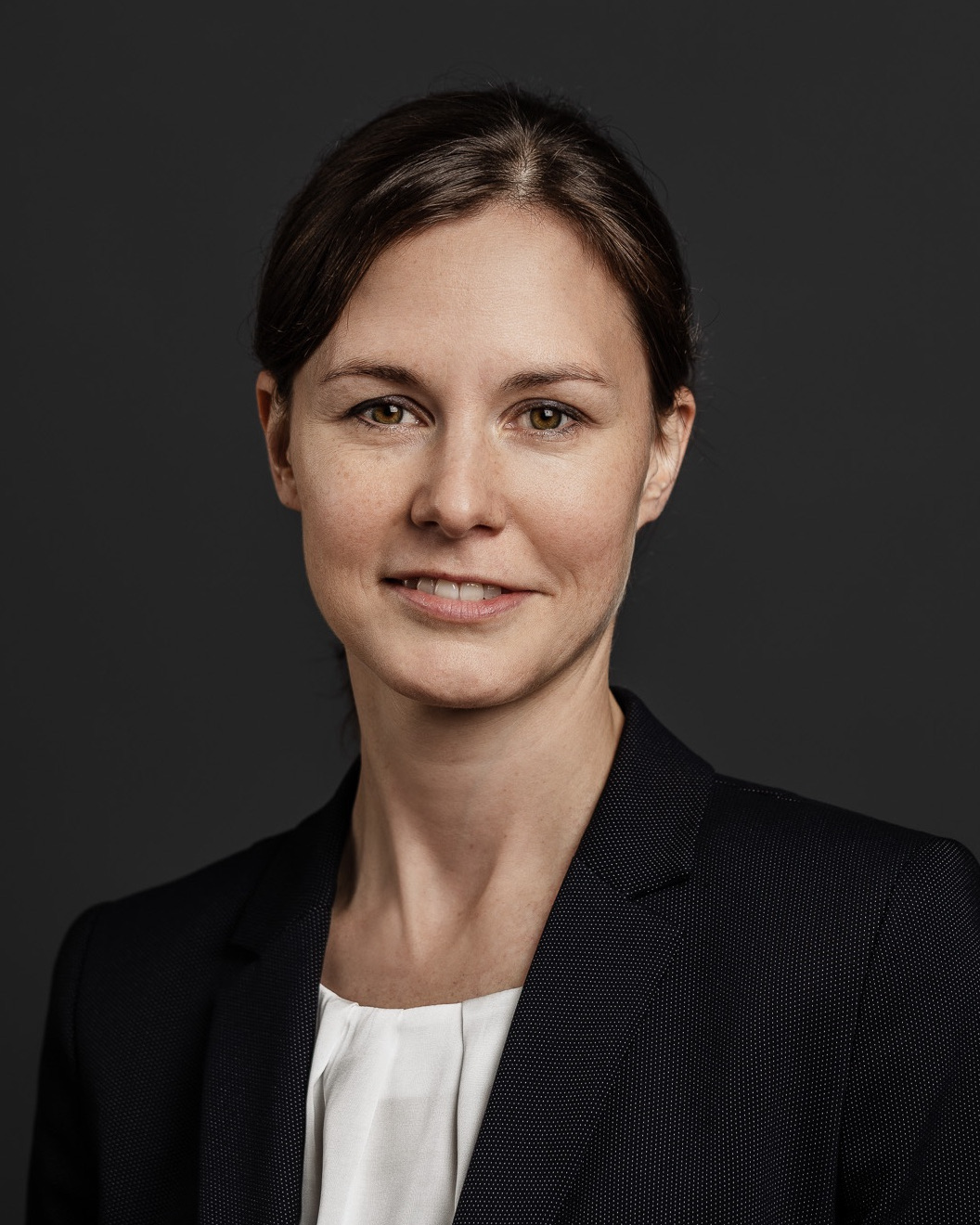}
}]{Melanie N. Zeilinger}
is an Associate Professor
at ETH Zürich, Switzerland. She received the
Diploma degree in engineering cybernetics from
the University of Stuttgart, Germany, in 2006,
and the Ph.D. degree with honors in electrical
engineering from ETH Zürich, Switzerland, in
2011. From 2011 to 2012 she was a Postdoctoral
Fellow with the École Polytechnique Fédérale de
Lausanne (EPFL), Switzerland. She was a Marie
Curie Fellow and Postdoctoral Researcher with
the Max Planck Institute for Intelligent Systems,
Tübingen, Germany until 2015 and with the Department of Electrical Engineering and Computer Sciences at the University of California at Berkeley, CA, USA, from 2012 to 2014. From 2018 to 2019 she was a professor at the University of Freiburg, Germany. Her current research interests include safe learning-based control, with applications to robotics and human-in-the loop control.
Dr. Zeilinger was the recipient of the ETH medal for her Ph.D. thesis, an SNF Professorship, the ETH Golden Owl for exceptional teaching in 2022, and the European Control Award in 2023.
\end{IEEEbiography}

\vfill

\end{document}

%% file: macros.tex
\newcommand{\state}[1][]{x_{#1}}
\newcommand{\optstate}[1][]{x_{#1}}
\newcommand{\optstateHat}[1][]{\hat{x}_{#1}}

\newcommand{\inputvar}[1][]{u_{#1}}
\newcommand{\inputvarHat}[1][]{\hat{u}_{#1}}
\newcommand{\mean}[2][]{\mu^{#1}_{#2}}
\newcommand{\meanHat}[2][]{\hat{\mu}^{#1}_{#2}}
\newcommand{\covar}[2][]{\Sigma^{#1}_{#2}}
\newcommand{\covarHat}[2][]{\hat{\Sigma}^{#1}_{#2}}

%% file: svg-inkscape/l4acados_flowchart_no_overlap_svg-tex.pdf_tex
\begingroup%
  \makeatletter%
  \providecommand\color[2][]{%
    \errmessage{(Inkscape) Color is used for the text in Inkscape, but the package 'color.sty' is not loaded}%
    \renewcommand\color[2][]{}%
  }%
  \providecommand\transparent[1]{%
    \errmessage{(Inkscape) Transparency is used (non-zero) for the text in Inkscape, but the package 'transparent.sty' is not loaded}%
    \renewcommand\transparent[1]{}%
  }%
  \providecommand\rotatebox[2]{#2}%
  \newcommand*\fsize{\dimexpr\f@size pt\relax}%
  \newcommand*\lineheight[1]{\fontsize{\fsize}{#1\fsize}\selectfont}%
  \ifx\svgwidth\undefined%
    \setlength{\unitlength}{440.09630098bp}%
    \ifx\svgscale\undefined%
      \relax%
    \else%
      \setlength{\unitlength}{\unitlength * \real{\svgscale}}%
    \fi%
  \else%
    \setlength{\unitlength}{\svgwidth}%
  \fi%
  \global\let\svgwidth\undefined%
  \global\let\svgscale\undefined%
  \makeatother%
  \begin{picture}(1,0.37531824)%
    \lineheight{1}%
    \setlength\tabcolsep{0pt}%
    \put(0,0){\includegraphics[width=\unitlength,page=1]{l4acados_flowchart_no_overlap_svg-tex.pdf}}%
    \put(0.62375115,0.08998689){\makebox(0,0)[t]{\lineheight{1.25}\smash{\begin{tabular}[t]{c}Preparation\end{tabular}}}}%
    \put(0.62315442,0.13015986){\makebox(0,0)[t]{\lineheight{1.25}\smash{\begin{tabular}[t]{c}\pyth{AcadosOcpSolver}\end{tabular}}}}%
    \put(0.89774971,0.08606294){\makebox(0,0)[t]{\lineheight{1.25}\smash{\begin{tabular}[t]{c}\pyth{AcadosOcp}~\eqref{eq:OCP_linear_model}\\(LTV dynamics)\end{tabular}}}}%
    \put(0.89526221,0.24757608){\makebox(0,0)[t]{\lineheight{1.25}\smash{\begin{tabular}[t]{c}\pyth{AcadosOcp}~\eqref{eq:OCP_GPMPC}\\(Nonlin. dynamics)\end{tabular}}}}%
    \put(0.62315441,0.30928889){\makebox(0,0)[t]{\lineheight{1.25}\smash{\begin{tabular}[t]{c}\pyth{AcadosSimSolver}\end{tabular}}}}%
    \put(0.1714801,0.21709419){\makebox(0,0)[t]{\lineheight{1.25}\smash{\begin{tabular}[t]{c}\pyth{ResidualModel}\end{tabular}}}}%
    \put(0.4363974,0.23536084){\makebox(0,0)[t]{\lineheight{1.25}\smash{\begin{tabular}[t]{c}\texttt{get}\end{tabular}}}}%
    \put(0.43271237,0.00904078){\makebox(0,0)[t]{\lineheight{1.25}\smash{\begin{tabular}[t]{c}\texttt{get} $\hat{x}_k, \hat{u}_k$\end{tabular}}}}%
    \put(0.43629219,0.28857628){\makebox(0,0)[t]{\lineheight{1.25}\smash{\begin{tabular}[t]{c}\texttt{set} $\hat{x}_k, \hat{u}_k$\end{tabular}}}}%
    \put(0.28291666,0.3243639){\color[rgb]{0,0.34117647,0.80784314}\makebox(0,0)[t]{\lineheight{1.25}\smash{\begin{tabular}[t]{c}provide\end{tabular}}}}%
    \put(0.84835971,0.15415953){\color[rgb]{0.8,0,0.19607843}\makebox(0,0)[t]{\lineheight{1.25}\smash{\begin{tabular}[t]{c}process\end{tabular}}}}%
    \put(0.43458833,0.11271685){\makebox(0,0)[t]{\lineheight{1.25}\smash{\begin{tabular}[t]{c}\texttt{set}\end{tabular}}}}%
    \put(0.17212871,0.16491222){\makebox(0,0)[t]{\lineheight{1.25}\smash{\begin{tabular}[t]{c}Compute\\$g, \frac{\partial g}{\partial x},\frac{\partial g}{\partial u}$\end{tabular}}}}%
    \put(0,0){\includegraphics[width=\unitlength,page=2]{l4acados_flowchart_no_overlap_svg-tex.pdf}}%
    \put(0.62469382,0.25922234){\makebox(0,0)[t]{\lineheight{1.25}\smash{\begin{tabular}[t]{c}Compute\\$f, \frac{\partial f}{\partial x},\frac{\partial f}{\partial u}$\end{tabular}}}}%
    \put(0,0){\includegraphics[width=\unitlength,page=3]{l4acados_flowchart_no_overlap_svg-tex.pdf}}%
    \put(0.62375115,0.03039411){\makebox(0,0)[t]{\lineheight{1.25}\smash{\begin{tabular}[t]{c}Feedback\end{tabular}}}}%
    \put(0,0){\includegraphics[width=\unitlength,page=4]{l4acados_flowchart_no_overlap_svg-tex.pdf}}%
    \put(0.43186528,0.06862711){\makebox(0,0)[t]{\lineheight{1.25}\smash{\begin{tabular}[t]{c}$\hat{A}_k, \hat{B}_k, \hat{c}_k$\end{tabular}}}}%
    \put(0.43629218,0.19143845){\makebox(0,0)[t]{\lineheight{1.25}\smash{\begin{tabular}[t]{c}$f, \frac{\partial f}{\partial x},\frac{\partial f}{\partial u}$\end{tabular}}}}%
  \end{picture}%
\endgroup%